\documentclass[11pt,preprint,nofootinbib]{revtex4-1}

\usepackage{graphicx}
\usepackage{latexsym}
\usepackage{epstopdf}
\usepackage{amssymb,amsfonts,amsmath}

\newcommand{\E}{\mathrm{E}}
\newcommand{\Var}{\mathrm{Var}}

\newcommand{\prob}{\mathcal{P}}
\newcommand{\D}{\langle d \rangle}
\newcommand{\pd}{\mathcal{D}}

\begin{document}

\title{The cause of universality in growth fluctuations}

\author{Yonathan Schwarzkopf$^{1,2}$, Robert L. Axtell$^{3,2}$ and J. Doyne Farmer$^{2,4}$} 
\affiliation{$^1$ California Institute of Technology, Pasadena, CA 91125 \\
$^2$ Santa Fe Institute, 1399 Hyde Park Road, Santa Fe, NM 87501\\
$^3$ George Mason University, 4400 University Drive, Fairfax, Virginia 22030\\
$^4$ LUISS Guido Carli, Viale Pola 12, 00198 Roma, Italy}


\begin{abstract}   
Phenomena as diverse as breeding bird populations, the size of U.S. firms, money invested in mutual funds, the GDP of individual countries and the scientific output of universities all show unusual but remarkably similar growth fluctuations. The fluctuations display characteristic features, including double exponential scaling in the body of the distribution and power law scaling of the standard deviation as a function of size.  To explain this we propose a remarkably simple additive replication model:  At each step each individual is replaced by a new number of individuals drawn from the same replication distribution.  If the replication distribution is sufficiently heavy tailed then the growth fluctuations are Levy distributed.  We analyze the data from bird populations, firms, and mutual funds and show that our predictions match the data well, in several respects: Our theory results in a much better collapse of the individual distributions onto a single curve and also correctly predicts the scaling of the standard deviation with size.  To illustrate how this can emerge from a collective microscopic dynamics we propose a model based on stochastic influence dynamics over a scale-free contact network and show that it produces results similar to those observed.  We also extend the model to deal with correlations between individual elements.  Our main conclusion is that the universality of growth fluctuations is driven by the additivity of growth processes and the action of the generalized central limit theorem.
 \end{abstract}

\maketitle

\section{Introduction}
 Recent research has revealed surprising properties in the fluctuations in the size of entities such as breeding bird populations along given migration routes \cite{Keitt98}, U.S. firm size \cite{Stanley96,Amaral97a,Bottazzi03a, Matia04,Bottazzi05,Axtell06}, money invested in mutual funds \cite{Schwarzkopf08},  GDP \cite{Canning98,Lee98,Castaldi09}, scientific output of universities \cite{Matia05}, and many other phenomena \cite{Plerou99,Keitt02,Bottazzi07,Podobnik08,Rozenfeld08}.  This is illustrated in Figures~1 and 2.  The first unusual property is in the logarithmic annual growth rates $g_t$, defined as  \mbox{$g_t=\log(N_{t+1}/N_{t})$}, where $N_t$ is the size in year $t$.  As seen in the top panel of Figure~1, all of the data sets show a similar double exponential scaling in the body of the distribution, indicating heavy tails.  The second surprising feature is the power law scaling of the standard deviation $\sigma$ with size, as illustrated in Figure~2.  In each case the standard deviation scales as $\sigma \sim N^{-\beta}$ with $\beta \approx 0.3$.
 
These results are viewed as interesting because they suggest a non-trivial collective phenomena with universal properties.  If the individual elements fluctuate independently, then (with a caveat we will state shortly) the standard deviation of the growth rates scales as a function of size with an exponent $\beta = 1/2$, whereas if the individual elements of the population move in tandem the standard deviation scales with $\beta = 0$, i.e. it is independent of size.  The fact that we instead observe a power law with an intermediate exponent $0 < \beta < 1/2$ suggests that the individual elements neither change independently nor in tandem.  Instead it suggests some form of nontrivial long-range coupling.  Why should phenomena as diverse as breeding bird populations and firm size show such similar behavior?  There is a substantial body of previous work attempting to explain individual pheomena, such as firm size or GDP \cite{Gibrat31,Defabritiis03,Fu05,Riccaboni08,Simon58,Ijiri75,Amaral97b,Buldyrev97,Amaral98,Bottazzi01,Sutton01,Wyart03,Bottazzi03b,Gabaix09,Schweiger07}.  However none of these theories has the generality to explain how this behavior could occur so widely.
 
The caveat in the above reasoning is the assumption that the fluctuations of the individual elements are well-behaved, in the sense that they are not too heavy-tailed.  As we show in a moment, if the growth fluctuations of the individual elements are sufficiently heavy-tailed then the fluctuations of the population are also heavy tailed, even if there are no collective dynamics.  Under the simple additive replication model that we propose the fluctuations in size are Levy distributed in the large $N$ limit.  This predicts a scaling exponent $0 < \beta < 1/2$ and the shape parameter of the Levy distribution predicts the value of $\beta$.  We show here that this model provides an excellent fit to the data.

In the first part of this paper we develop the additive replication model and show that it gives a good fit to the data.  Our analysis in the first part is predicated on the existence of a heavy-tailed replication distribution.  In the second part of the paper we present one possible explanation for the heavy-tailed replication distribution in terms of stochastic influence dynamics on a scale-free contact network, and argue that such an explanation could apply to any of the diverse settings in which these scaling phenomena have occurred.    This influence dynamics is an example of ``nontrivial" collective dynamics.   Thus, the process that generates the heavy tails in individual fluctuations may come from nontrivial collective dynamics even though the replication model does not depend on this.

 \section{The additive replication model}

\begin{table*}
\caption{\label{table_levy_fit}%
The parameter values for fitting the data with a Levy distribution }
\begin{tabular}{c|c|c|c|c}
\hline
year&$\alpha$&$\kappa$&c&$\mu$\\
\hline\hline
NABB&1.40&0.81&0.156&-0.037\\
\hline
Mutual funds&1.48&0.3&0.111&-0.015\\
\hline
Firms&1.53&0.80&0.16&-0.05\\
\hline
\end{tabular}
\end{table*}
%

\begin{figure}
\begin{center}
\includegraphics[width=9cm]{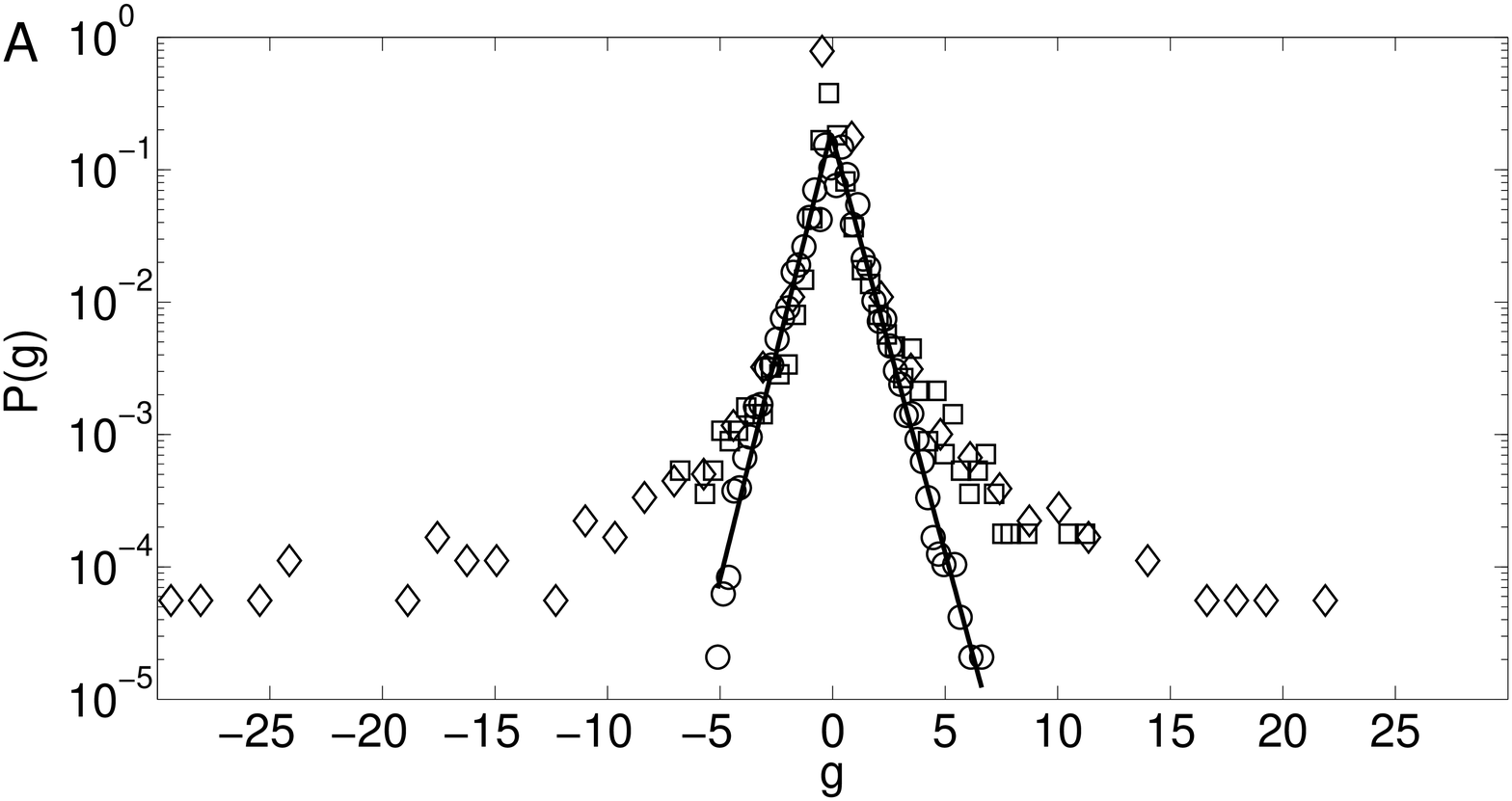}
\includegraphics[width=9cm]{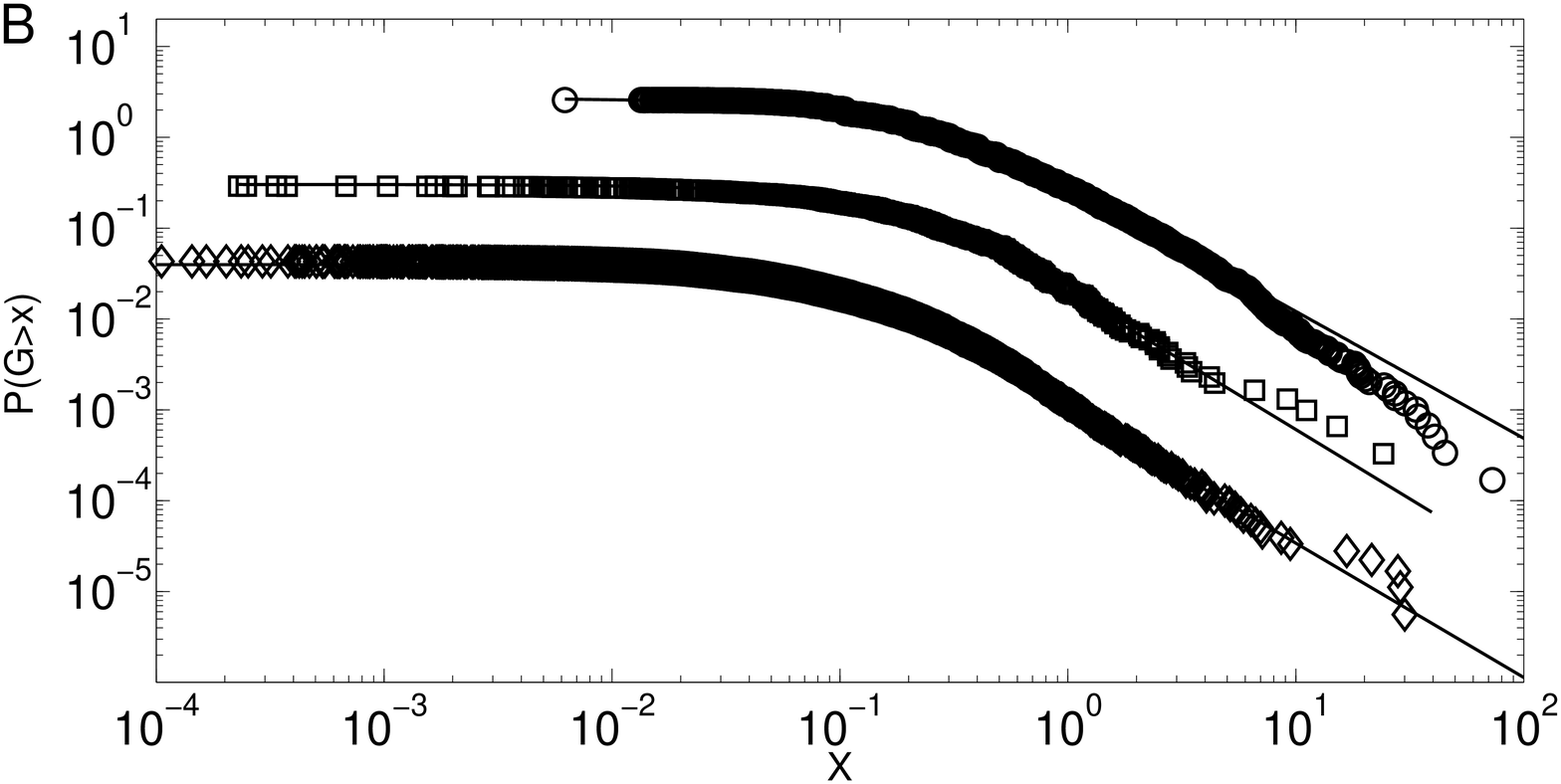}
\includegraphics[width=9cm]{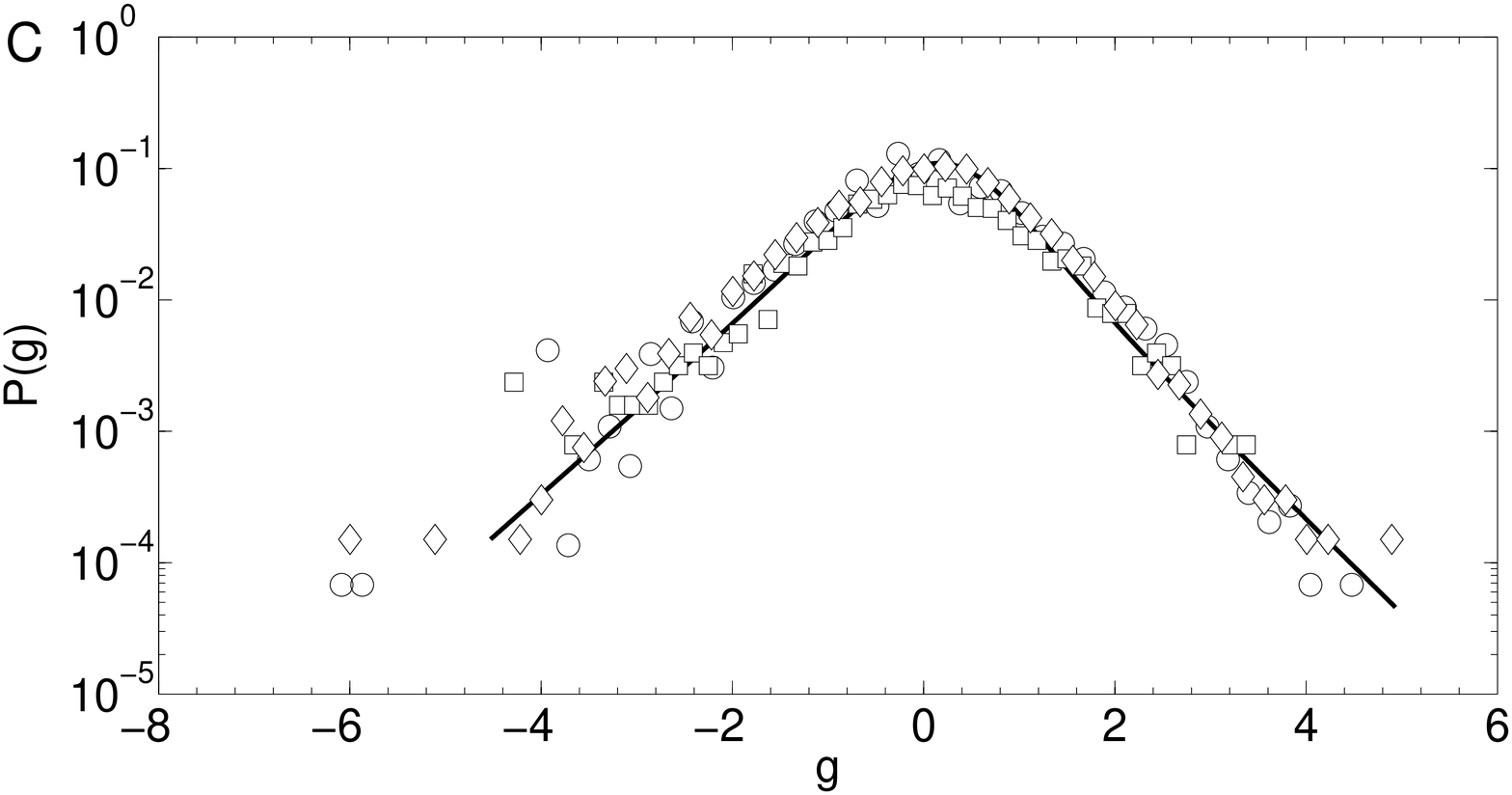}
\caption{ \footnotesize An illustration of how our theory reveals the underlying regularity in the distribution of growth fluctuations of highly diverse phenomena.  The three data sets studied here are North American Breeding Birds ($\circ$), US firm sales ($\Box$) and US equity mutual funds ($\diamondsuit$).  The data is the same in all three panels, the only change is the presentation.  {\bf A}: {\it The traditional view.}  Histograms of the logarithmic growth rates are plotted on semi-log scale, normalized such that the mean vanishes $\E[g]=0$ and the variance is unity $\Var[g]=1$.  The collapse is good for the body of the distribution, revealing double-exponential scaling, but poor in the tails, where the three data sets look quite different.
{\bf B}: {\it Comparison to a Levy distribution.}  The cumulative distribution $P(G>X)$ of relative growth rates for the three data sets are compared to fits to the Levy distributions predicted by our theory  (solid curves) and plotted on double logarithmic scale (for positive $X$ only). See table~2 for parameter values.  {\bf C}:  {\it Superior collapse onto a single curve when the data is scaled as predicted by our theory.}  The empirical values of the relative growth $G$ (rather than the logarithmic growth rate $g$) are normalized so they all have a scale parameter approximately one, as described in the text.  In order to compare to the top panel, we plot the logarithmic growth $g$ and compare to a Levy distribution (solid curve).  This gives a better collapse of the data which works in the tails as well as the body.}
\label{PDF}
\end{center}
\end{figure}

We assume an additive replication process: At each time step each individual element is replaced by $k$ new elements drawn at random from a {\it replication distribution} $p(k)$, where $0 \le k < \infty$.  An individual element could be a bird, a sale by a given firm, or the holdings of a given investor in a mutual fund.  By definition the number of elements $N_{t+1}$ on the next time step is
 \begin{equation}
 N_{t+1}=\sum_{j=1}^{N_t}k_{jt},
 \label{replicationModel}
 \end{equation}
where $k_{jt}$ is the number of new elements replacing element $j$ at time $t$.  
The growth $G_t$ is given by
\begin{equation}\label{Gt}
G_t=\frac{N_{t+1}-N_t}{N_t}=\frac{\sum_{j=1}^{N_t}k_{jt} }{N_t}-1.
\end{equation}
The simplest version of our model assumes that draws from the replication distribution $p(k)$ are independent; we later relax this assumption to allow for correlations.  

Why might such a model be justified? First note that additivity of the elements is automatic, since by definition the size is the sum of the number of elements.   The assumption that each element replicates itself in the next year amounts to a persistence assumption, i.e. that the number of elements in one year is linearly related to the number in the previous year, with each element influencing the next year independently of the others.  We also assume uniformity by letting all elements have the same replication distribution $p(k)$.  For the case of firms, for example, each sale in year $t$ can be viewed as replicating itself in year $t+1$.  This is plausible if the typical customer remains faithful to the same firm, normally continuing to buy the product from the same company, but occasionally changing to buy more or less of the product.  For migrating birds this is plausible if the number of birds taking a given route in a given year is related to the number taking it last year, either because of the survival probability of individual birds or flocks of birds, or because individual birds influence other birds to take a given migration route.

\section{Predictions of the model}

Given that the size $N_t$ at time $t$ is known and the drawings from $p(k)$ are independent, the growth rate $G_t$ is a sum of $N_t$ I.I.D. random variables.  Under the generalized central limit theorem \cite{Zolotarev86,Resnick07}, in the large $N_t$ limit the growth $\prob_G$ converges to a Levy skew alpha-stable distribution 
\begin{equation}\label{levy_scaling}
\prob_G(G_{t}|N_t)= N_t^{-\frac{1-\alpha}{\alpha}}L_{\alpha}^{\kappa}(G_t\,N_t^{-\frac{1-\alpha}{\alpha}};c,\mu).
\end{equation} 
$0<\alpha\leq 2$ is the shape parameter, $-1\leq\kappa\leq1$ is the asymmetry parameter, $\mu$ is the shift parameter and $c$ is a scale parameter.

The normal distribution is a special case corresponding to $\alpha=2$.  This occurs if the second moment of $p(k)$ is finite.  However, if the second moment diverges according to extreme value theory, under conditions that are usually satisfied, it is possible to write $p(k)\sim k^{-\gamma}$  for large $k$ \footnote{Under extreme value theory there are distributions for which there is no convergent behavior; the power law assumes convergence.}.  When $1 < \gamma < 3$  the Levy distribution has heavy tails that asymptotically scale as a power law with $P(G > x) \sim x^{-\alpha}$, where $\alpha=\gamma-1$.

The additive replication process theory predicts power law behavior for $\sigma(N)$ and predicts its scaling exponent based on the growth distribution.   If $\gamma>3$ the growth rate distribution converges to a normal with $\beta=1/2$ \,\footnote{%
 For $\gamma=3$ and $\gamma=2$ there are logarithmic corrections to the results.}.   However, when $\alpha = \gamma + 1 < 2$, using standard results in extreme value theory \cite{Zolotarev86,Resnick07} the standard deviation scales as a power law with size, $\sigma_G \sim N_t^{-\beta}$, where
 \begin{equation}
 \label{betaEQ}
 \beta=(\gamma-2)/(\gamma-1).
 \end{equation}
 
 \section{Testing the predictions}

To test the prediction that the data is Levy-distributed, in the central panel of Figure~1 
we compare each of our three data sets to Levy distributions.  The three data sets are  (1) the number of birds of a given species observed along a given migration route, (2) the size of a firm as represented by its sales, and (3) the size of a U.S. mutual fund.  The data shown in the middle panel of Figure 1 are exactly the same as in the upper panel, except that we plot the growth fluctuations $G$ rather than their logarithmic counterpart $g$, we plot a cumulative distribution rather than a histogram, and we graph the data on double logarithmic scale.  The fits are all good.

Because we are lucky enough that the shape parameter $\alpha$ and the asymmetry parameter $\kappa$ are similar in all three data sets, we can collapse them onto a single curve. This is done by transforming all the data sets to the same scale in $G$ by dividing by an empirically computed scale factor equal to the 0.75 quantile minus the 0.25 quantile (we do it this way rather than dividing by the standard deviation because the standard deviation does not exist).  It is important that this normalization is done in terms of $G$, in contrast to the standard method which normalizes the logarithmic growth $g$.  The standard method, illustrated in the top panel, produces a collapse for the body of the distribution, but there is no collapse for the tails --  mutual funds have very heavy tails while the breeding birds closely follow the exponential even for large values of $g$.  In contrast, the collapse using $G$ as suggested by our theory, illustrated in the bottom panel, works for both the body and the tails.

\begin{figure}
\begin{center}
\includegraphics[width=9cm]{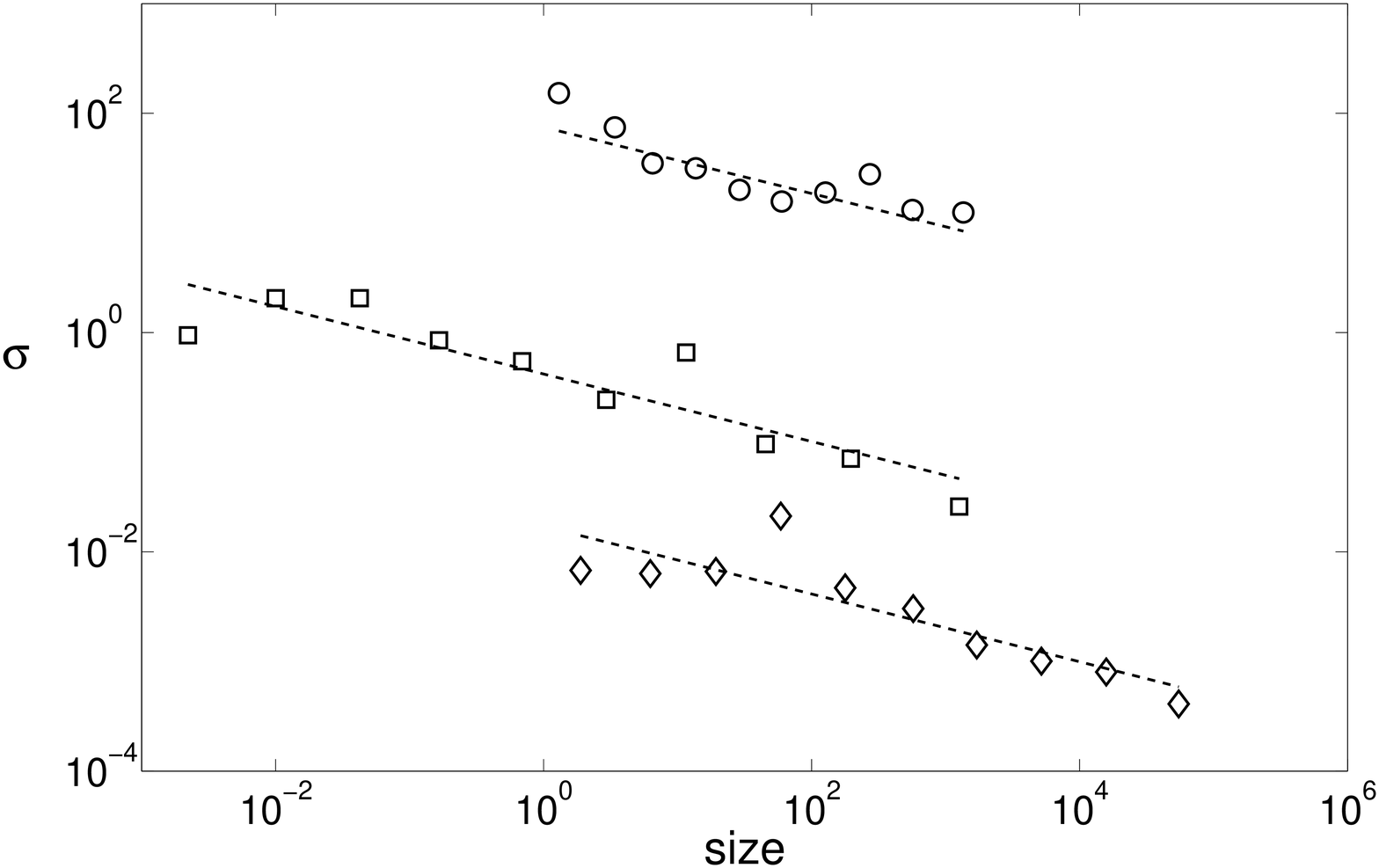}
\caption{Illustration of the non-trivial scaling of the standard deviation $\sigma$ as a function of size $N$.  The straight lines on double logarithmic scale indicate power law scaling.  Same symbols as in Figure~1. 
The standard deviation is computed by binning the data into bins of exponentially increasing size and computing the sample standard deviation in each bin.
For clarity the breeding bird population is shifted by a factor of $10$ and the mutual fund data set by a factor of $10^{-1}$.  The empirical data are compared to lines of slopes $-0.303$, $-0.308$ and $-0.309$ respectively.}
\label{scaling}
\end{center}
\end{figure}

\begin{table}
\caption{\label{table_fit}%
A demonstration that the Levy distribution makes a good prediction of the scaling of the standard deviation as a function of size.  The measured value of $\gamma$ based on the center panel of Figure~1  is used to make a prediction, $\hat{\beta}$, of the exponent of the scaling of the standard deviation.  This is in good statistical agreement with $\beta$, the measured value.  NABB stands for North American Breeding Birds.}
\begin{tabular}{c|c|c|c}
\hline
year&$\beta$&$\hat{\beta}$&$\gamma$\\
\hline\hline
NABB&$0.30\pm0.07$&$0.29\pm0.03$&$2.40\pm0.06$\\
\hline
Mutual funds&$0.29\pm0.03$&$0.32\pm0.04$&$2.48\pm0.08$\\
\hline
Firms&$0.31\pm0.07$&$0.35\pm0.03$&$2.53\pm0.07$\\
\hline
\end{tabular}
\end{table}

To test the prediction of the power law scaling of the standard deviation with size we estimated $\gamma$ from the data shown in Figure~1 and $\beta$ from the data in Figure~2. 
We then make a prediction $\hat{\beta}$ for each data set using Eq.~\ref{betaEQ} and the estimated value of $\gamma$ for each data set. 
The results  given in Table~1 
are in good statistical agreement in every case.  (See Materials and Methods.)

\section{Why is the replication distribution heavy-tailed?}

Part of the original motivation for the interest in the non-normal properties and power law scalings of the growth fluctuations is the possibility that they illustrate an interesting collective growth phenomenon with universal applicability ranging from biology to economics.  Our explanation so far seems to suggest the opposite:  In our additive replication model each element acts independently of the others.  As long as the replicating distribution is heavy tailed the scaling properties illustrated in Figures~1 and 2 will be observed, even without any collective interactions.

There is a subtle point here, however.  Our discussion so far leaves open the question of why the replication distribution might be heavy-tailed.  Based on the limited data that is currently available there are many possible explanations -- it is not possible to choose one over another.   One can postulate mechanisms that involve no collective behavior at all, for example, if individual birds had huge variations in the number of surviving offspring.  (This might be plausible for mosquitos but does not seem plausible for birds).  One can also postulate mechanisms that involve collective behavior, as we do in the next section.

\section{The contact network explanation for heavy tails}

In this section we present a plausible explanation for power law tails of $p(k)$ in terms of random influence on a scale-free contact network.  This example nicely illustrates how the heavy tails of the individual replication distribution $p(k)$ can be caused by a collective phenomenon.

Assume a contact network \cite{Dorogovtsev08} where each node represents individuals.  They are connected by an edge if they influence each other.  For simplicity assume that influence is bi-directional and equal, i.e. that the edges are undirected and unweighted.    Let individual $i$ be connected to $d_i$ other individuals, where $d_i\in\{1,\ldots M\}$ is the degree of the node.  The degree distribution $\pd(d)$ is the probability that a randomly selected node has degree $d$.

\begin{figure}
\begin{center}
\includegraphics[width=6cm]{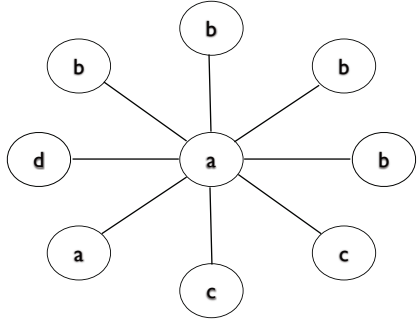}
\caption{\label{graph_example}
Here we show an example of a simple network. Each node represents an individual and each edge represents a contact between them. The labels
represent the group the individual belongs to.}
\end{center}
\end{figure}

Let each individual belong to one of $\Gamma$ groups. For example, belonging to group $a\in\{1\ldots\Gamma\}$ can represent a consumer owning a product of firm $a$, an investor with money in mutual fund $a$, or a bird of a given species taking migration route $a$. The groups are the same as the populations discussed earlier, i.e. $N^a_t$ is the size of group $a$ at time $t$. The dynamics are epidemiological in the sense that an individual will stay in her group unless her contacts influence her to switch. The switching is stochastic: An individual in group $a$ with a contact in group $b$ will switch to group $b$  with a rate $\rho_{ab}$. Furthermore,  the switching rate is linearly proportional to the number of contacts in that group, i.e. if an individual belonging to group $a$  has $n$ contacts in group $b$, she will switch with a rate $n\rho_{ab}$. As an example, the individual in the center of the graph in Fig~\ref{graph_example} has a degree $d=8$ and belongs to group $a$. She will switch to group $b$ with a rate $4\rho_{ab}$, to group $c$ with a rate $2\rho_{ac}$ and to group d with a rate $\rho_{ad}$.  
      
For example consider firm sales.  If a given consumer likes the product of a given firm, she might influence her friends to buy more, and if she doesn't like it, she might influence them to buy less.  Thus each sale in a given year influences the sales in the following year.  A similar explanation applies to mutual funds, under the assumption that each investor influences her  friends, or it applies to birds, under the assumption that each bird influences other birds that it comes into contact with\footnote{
It has recently been shown that influence in flocking pigeons is hierarchical. \cite{Kurvers09,Nagy10}.}.
%

We now show how the contact network gives rise to an additive replication model.  To calculate $N^a_{t+1}$ consider each of the $N^a_t$ individuals in group $a$ one at a time.  Individual $j$ in group $a$ replicates if she remains in the group, and/or if one or more of her contacts that belong to other groups join group $a$.  She fails to replicate if she leaves the group and also fails to influence anyone else to join.  Let the resulting number of individuals that replace individual $j$ be $k_{jt}$.  This implies
\begin{equation}\label{net_replication}
N^a_{t+1}=\sum_{j\in\mathrm{Group\,\, a}}^{N^a_t}k_{jt},
\end{equation}
which is identical to Eq.~\ref{replicationModel} except for the group label (which was previously implicit).

The replication factor $k_{jt}$ is a random number with values in the range $k_i\in[0,d_j]$. 
Given the stochastic nature of the influence process we approximate\footnote{
This approximation is valid  for random networks, which have a local tree-like structure \cite{Dorogovtsev08}.}
$k_{jt}$ as a Poisson random variable with mean $E[k_{jt}]=(1-\theta_a)d_j$, where $\theta_a$ is the probability that a randomly selected contact belongs to group $a$.     This means that the replication factor $k_{jt}$ is proportional to the degree, i.e. $k_{jt} \sim d_j$, and that the replication distribution is proportional to the degree distribution,
\begin{equation}
p(k)\approx (1-\theta_a) \pd((1-\theta_a) d).
\end{equation}

Thus the influence dynamics of the contact network are an additive replication process with the individual replication distribution proportional to the degree distribution of the network.  If the network is scale free, i.e. if for large $k$ the degree distribution is a power law with $\gamma < 3$, then the growth fluctuations will be Levy distributed.  It is beyond the scope of this paper to explain why the contact groups in the various settings that have been studied might be scale free, but there is at this point a large literature demonstrating that such behavior is common \cite{Albert02,Newman03}.

\begin{figure}
\begin{center}
\includegraphics[width=9cm]{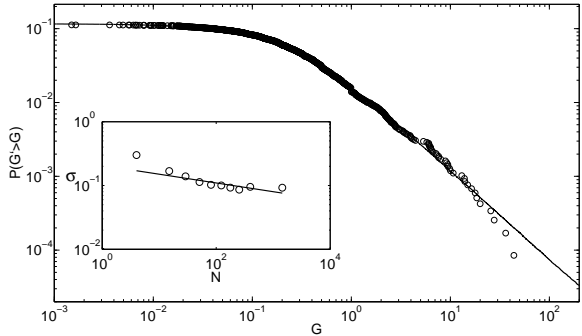}
\caption{\label{network_results}%
A demonstration that influence dynamics on a scale-free contact network give rise to the Levy behavior predicted by the additive replication model.  The influence model was simulated for $10^3$ groups on a network of $10^6$ nodes, an average degree $\D=10$ and a power law  degree distribution  $\pd(d)\sim d^{-\gamma}$ with $\gamma=2.2$.
The cumulative growth rate distribution $P(G'>G)$ is in good agreement with the predicted Levy distribution (\ref{levy_scaling}) 
{\bf Inset:} the fluctuations are compared to a line of slope $\beta=-0.1667$, illustrating the expected power law scaling. }
\end{center}
\end{figure}

A numerical simulation verifies these results\footnote{
The average number of individuals and the average growth rate of a group can be approximated using a mean field approach.  The mean field growth rates are given by
$\partial N_a/\partial t= \D M \theta_a (1-\theta_a)\sum_{b=1}^\Gamma (\rho_{ab}-\rho_{ba})$\cite{PastorSatorras01}.
and  \mbox{$ \theta_a=\D^{-1}\sum_{d'} d'\, f^{d'}_a \pd(d')$},
where $f^d_a$ is the fraction of individual elements with degree $d$ that belong to group $a$.  We know of no analytic method to compute the growth fluctuations.}. 
We simulated a network of $10^6$ nodes with a power law tailed degree distribution $\pd(d)\sim d^{-\gamma}$ with $\gamma=2.2$ and average degree $\D=10$. The dynamics were simulated for $10^3$ groups with a homogeneous switching rate $\rho_{ab}=\rho$.   As expected the growth rates have a Levy distribution $P(G)\sim G^{-\gamma}$ as shown in Figure~\ref{network_results}. The fitted parameter values are $\alpha=1.2$, $\kappa=0.25$, $c=0.09$ and $\mu=-0.17$. 
 The fitted value of the fluctuation scaling  $\beta=0.14\pm0.03$, shown in the inset of Figure~\ref{network_results}, is in agreement with the predicted value of $\beta=(\gamma-2)/(\gamma-1)=1/6$.    

\section{Correlations and finite size effects}

So far we have assumed that the growth process for individual elements is uncorrelated, i.e. that the draws from $p(k)$ are I.I.D.  Sufficiently strong correlations can change the results substantially.  There can be correlations among the individual elements or correlations in time.  For example, suppose some groups are intrinsically more or less popular than others.  For example, the popularity of a city might depend on its economy and living conditions.  This can be modeled by assuming that the replication of individual $j$ in group $i$ is given by a random variable $\hat{k}^i_{jt}$ which is the sum of a random variable that depends on the individual and one that is common for the group, i.e.  $\hat{k}^i_{jt} = k_{jt} + \zeta_{it}$.  We can then write the replication model in the form
\begin{equation}
N_{t+1}=  \sum_{i = 1, j=1}^{\Gamma, N^i_t}k_{jt} + \zeta_{it}.
 \label{corrRepModel}
 \end{equation}

As shown in the supplementary materials, for small sizes the individual fluctuations $k_{jt}$ dominate, so that there is a power law scaling of $\sigma$, but for larger sizes the group fluctuations $\zeta_{it}$ dominate, and $\sigma$ becomes constant (i.e. $\beta = 0$).  
 This is indeed what we observe for cities\footnote{
Note that the nature of the scalings for cities is controversial and strongly depends on how a city is defined -- our results are in agreement with those who claim the scaling is not very good \cite{Eeckhout04}. Rather than using the census definitions, Rozenfeld et al \cite{Rozenfeld08} use a clustering algorithm for defining cities and then the fluctuation scaling (without the group correlations) seems to hold.}.

We have also assumed in our analysis that the number of elements is infinite, i.e. that there is no upper limit on the replication factors. For finite systems the growth of one group is at the expense of another. This can induce correlations which affect both the growth rate distribution and the fluctuation scaling.  Nevertheless, as our simulation shows, under appropriate circumstances the theory can still describe finite systems to a very good approximation.  A more detailed discussion is provided in the supplementary  materials. 

\section{Discussion}

The explanation that we offer here is widely applicable and very robust.  The idea that a larger entity can be decomposed into a sum of smaller elements, and that the smaller elements can be modeled as if they replicate, is quite generic.  As discussed in the previous section this can be broken if the growth of the elements is too correlated.  
Our explanation for the heavy tailed growth rate distributions and fluctuation scaling requires that the replication distribution $p(k)$ is heavy tailed.  The key thing we have shown is that when this occurs, the generalized central limit theorem dictates that the growth distribution $P_G$ will be Levy, which in turn dictates the power law size dependence of the standard deviation, $\sigma(N)$. 

The previous models which are closest to ours are the model of firm size of Wyart and Bouchaud \cite{Wyart03} and the model of GDP due to Gabaix \cite{Gabaix09}.   Both of these models assume that the  size distribution $P(N_t)$ has power law tails and that firms grow via multiplicative fluctuations.  They each suggested (without any testing) that additivity might lead to Levy distributions for their specific phenomena (GDP or firm size).  This is in contrast to our model, which requires neither the assumption of power tails for size nor multiplicative growth.  This is a critical point because the size of mutual funds does not obey a power law distribution \cite{Schwarzkopf08}, which rules out both the the Wyart and Bouchaud and Gabaix models as general explanations.  We are apparently the first to realize that these diverse phenomena all obey Levy distributions, and that this explains the power law scaling of $\sigma(N)$.

There are many possible explanations that could generate a heavy tailed replication distribution $p(k)$.  Here we proposed an influence process on a scale free contact network as a possible example.  This mechanism is quite general and relies on the assumption that an individual element's actions are affected by those of its contacts.  Scale free networks are surprisingly ubiquitous and the existence of social, information and biological networks with power law tails with $2 < \gamma < 3$ is well documented \cite{Albert02,Newman03}, and suggests that the assumption that the degree distribution $\mathcal{D}(d)$ and hence the replication distribution $p(k)$ are heavy-tailed is plausible.

The influence model shows that the question of whether the interesting scaling properties of these systems should be regarded as ``interesting collective dynamics" can be subtle.  On one hand the additive replication model suppresses this -- any possibility for collective action is swept into the individual replication process.  On the other hand, the influence model shows that the heavy tails may nonetheless come from a collective interaction.  More detailed data is needed to make this distinction.

Our model shows that, whenever its assumptions are satisfied, one should expect universal behavior as dictated by the central limit theorem:  The growth fluctuations should be Levy distributed (with the normal distribution as a special case).  Our model does not suggest that the tail parameter should be universal, though of course this could be possible for other reasons. Based on our model there is no reason to expect that the value of the exponent $\alpha$ (or equivalently $\gamma$ or $\beta$) will not depend on factors that vary from example to example.  Thus the growth process is universal in one sense but not in another.

\section{Materials and methods}
\subsubsection{North american breeding birds dataset}
We use the the North American breeding bird survey, 
which contains 42 yearly observations for over 600 species along more than 3,000 observation routes.
 For each route the number of birds from each species is quoted for each year in the period 1966-2007. 
For each year in the data set, from 1966 to 2007, we computed the yearly growth with respect to each species in each route.
 The data set can be found online at \\
ftp://ftpext.usgs.gov/pub/er/md/laurel/BBS/DataFiles/.
\\
\subsubsection{US public firms dataset}
We use the 2008 COMPUSTAT dataset containing information on all US public firms. 
As the size of a firm we use the dollar amount of sales.  Growth is given by the 3 year growth in sales. 
\subsubsection{US equity mutual fund dataset}
We use the Center for Research in Security Prices (CRSP) mutual fund database, 
restricted to equity mutual funds existing in the years 1997 to 2007.
An equity fund is one with at least 80\% of its portfolio in stocks.
As the size of the Mutual fund we use the total net assets value (TNA) in real US dollars as  reported monthly.  
Growth in the mutual fund industry, measured by change in TNA, is comprised of two sources: growth due to the funds performance and growth due to flux of money from investors, i.e. mutual funds can grow in size if their assets increase in value or due to new money coming in from investors.
We define the relative growth in the size of a fund at time $t$ as 
\[G_{TNA}(t)=\frac{TNA_{t+1}}{TNA_t}-1\]
and decompose it as follows;
\begin{equation}
G_{TNA}(t)=r_t+G_t,
\end{equation} 
where $r_t$ is the fund's return, quoted monthly in the database, and $G_t$ is the growth due to investors.  
For our purposes here we only consider $G_t$, the growth due to investors. 

\subsection{Empirical fitting procedures}
The empirical investigation is conducted as follows:  We first estimate the fluctuation scaling exponent $\beta$.   The relative growth rate distribution $G=N_{t+1}/N_t-1$ is binned into 10 exponentially spaced bins according to size $N_t$. For each bin $i$, the sample estimate of the variance of the growth rates  $\sigma_i^2$ is estimated in the usual way.
Then the logarithm of the measured variances are regressed on the logarithm of the average size $\bar{N}_i$
\begin{equation}
\log(\sigma)=\beta \log(N)+\sigma_1
\end{equation}
 such that the slope is the ordinary least squares (OLS) estimator of $\beta$.

To estimate the tail exponent we normalize the growth rate $G$ such that it has zero mean and we divide by the 0.75 quartile - the 0.25 quartile.  We estimate tail exponents using the technique described in Clauset et al \cite{Clauset07}. The method used uses the following modified Kolmogorov-Smirnoff statistic 
\[KS=\max_{x>x_{min}}\frac{|s(x)-p(x)|}{\sqrt{p(x)[1-p(x)]}},\] 
where $s$ is the empirical cumulative distribution and $p$ is the hypothesized cumulative distribution.  Using the maximum likelihood estimator (MLE) of the tail exponent $\gamma$  we can predict the fluctuation scaling exponent $\hat{\beta}$ using Eq.~4. and compare to the measured OLS estimator of $\beta$.

\begin{acknowledgments}
We gratefully acknowledge financial support from NSF grant HSD-0624351.
\end{acknowledgments}



\end{document}


\author{Yonathan Schwarzkopf, Robert L. Axtell and J. Doyne Farmer}
\title{SUPPORTING INFORMATION\\{ The cause of universality in growth fluctuations}}

\maketitle

As supplementary materials we provide the following:
In Section~\ref{sec:constraint} we discuss a modification to our model incorporating a constraint over the number of elements. 
In Section~\ref{sec:log_growth_Rate} we derive and discuss in detail the logarithmic growth rate distribution arising from our model. In Section~\ref{sec:correlations} we discuss a modification to the model that incorporates correlations in the replication factors of elements belonging to the same group.  
In Section~\ref{sec:empirics} we describe in more detail the fitting procedures and the data sets used in our empirical investigation. We provide in this section additional analysis supporting our results. 
 \section{Growth under a constrained number of elements}\label{sec:constraint}

Our model, as defined in the manuscript, assumes that there are infinitely many elements. We examine here a setting for growth in which several groups compete over a fixed number of elements $N_{total}$. In this setting elements that join one group must do so at the expense of another group resulting in a decrease in the group size. 
 At each time step the number of elements in group $a$ changes according to the replication process such that
 \begin{equation}
 N^a_{t+1}=\sum_{j=1}^{N^a_t}k_{jt},
 \label{replicationModel}
 \end{equation}
where $k_{jt}$ is the number of new elements replacing element $j$ at time $t$.   
The net change in the number of elements in group $a$ is defined as
 \[\Delta^a_t=N^a_{t+1}-N^a_t\]
 drawn from the distribution $ P_\Delta(|\Delta_t|\,|N_t)$.
 This distribution is equivalent to the distribution of $N_{t+1}$, a distribution of a sum of $N_t$ random variables, described in the main text.
For more than a single group, under  the constraint on the number of elements,  a negative net size change in a group correspond to elements leaving the group to join other groups. 
We generalize the distribution of the net size change in group $a$ as $ \prob_\Delta(|\Delta^a_{t}|\,|N^a_{t})$  as 
\begin{eqnarray}\label{P_delta}
\prob_\Delta(\Delta_{it}|N_t)&=&P_\Delta(|\Delta_{it}|\,|N_t)\Theta(\Delta_{it})\Theta(\NT-N_{it}-\Delta_{it})\\
&&+\left[\sum'_{\{\Delta^b_{t}\}}\sum'_{\{N^b_{j}\}}P_\Delta(|\Delta^b_{t}|\,|N^b_{t})\right]\Theta(-\Delta^a_{t})\Theta(N^a_{t}-|\Delta^a_{t}|),\nonumber
\end{eqnarray}
where $\Delta^b_{t}$ is the change in the number of elements in group $b$, $N^b_{t}$ is the occupation of that group and $\Theta$ is the unit step function. The summations are over all configurations where$\Delta^b_{t}\geq0$ and $N^b_{t}\geq0$ such that $\sum_{b\neq a} \Delta^b_{t}=\Delta^a_{t}$ and $\sum_{b\neq a} N^b_{t}=\NT-N_{it}$.
The first term in (\ref{P_delta}) corresponds to elements joining group $a$ and that number is bound from above by the number of elements in the other groups $\NT-N^a_{t}$. The second term corresponds to elements leaving group $a$ and is written in terms of elements joining the other groups.

Since the relative size change is given by $G_t=\Delta_t/N_t$ we can write the relative size change distribution as
$P_G(G|N)= N \prob_\Delta(N G |N)$ yielding 
\begin{eqnarray}\label{PG_constraint}
P_G(G^a_t|N^a_t)&=&N^a_t P_\Delta(|N^a_tG^a_t|\,|N^a_t)\Theta(G^a_t)\Theta(\frac{\NT-N^a_t}{N^a_t}-G^a_t)\\
&&+\left[\sum'_{\{G^b_t\}}\sum'_{\{N^b_t\}} N^b_t P_\Delta(|N^b_t G^b_t |\,|N^b_t)\right]\Theta(-G^a_t)\Theta(1-|G^a_t|) \nonumber
\end{eqnarray}
where the $G^b_t$ are constrained such that $G^b_t\geq-1$ and $\sum_b N^b_t G^b_t=N^a_tG^a_t$. 
The second term corresponding to elements leaving group $a$ is a sum of elements joining the different groups and as such is a sum of random variables and will depend on the distribution of the number of elements in a group.

The distribution can be written explicitly for $\Gamma \gg1$ groups with an equal number of elements $N=\NT/\Gamma\gg1$ for which  
\begin{eqnarray}
P_G(G|N)&=& \frac{1}{N^{1/\alpha-1}}L_{\alpha}^{\kappa}(\frac{G}{N^{1/\alpha-1}};c,\mu)\Theta(G)\Theta(\frac{\NT-N}{N}-G)\\
&&+ \frac{1}{\Gamma^{1/\alpha}N^{1/\alpha-1}}L_{\alpha}^{\kappa}(\frac{G}{\Gamma^{1/\alpha}N^{1/\alpha-1}};c,\mu)\Theta(-G)\Theta(1-|G|) \nonumber
\end{eqnarray}
In general, the growth distribution under a constraint on the number of elements (\ref{PG_constraint}) can be different than that  of the non constrained case described in the main text.

\section{The logarithmic growth rate distribution}\label{sec:log_growth_Rate}
The logarithmic size change $g=\log(N_{t+1}/N_t)$, as opposed to the relative size change $G$, is not explicitly modeled in our theory. However, it is used by  most researchers for its additivity property, which makes calculations easier and in order to compare this work with other models we describe here shortly the distribution of log changes $P_g(g)$ and its size scaling.

Given that we know the growth rate distribution $P_G$, the distribution of log size changes  can be achieved through a change of variables $G=\exp(g)-1$ which yields\footnote{ 
Under a change of variables the distribution functions transform such that $P_g(g)\mathrm{d}g=P_G(G)\mathrm{d}G$ holds. 
Given that we know $P_G(G)$ and a change of variables $G=f(g)$ the distribution $P_g(g)$ is then given by 
$P_g(g)=f'(g)P_G(f(g))$.}
\begin{equation}
P_g(g|N)=\e^{g} P_G(\e^{g}-1|N)
\end{equation} 
 For distribution $P_G$ that decay slow enough as $G$ approaches$-1$, i.e. in the limit $g\to -\infty$, the lower tail of the distribution  
 can be approximated as
 \begin{equation}\label{lower_tail}
 P_g(g\ll0)\sim\e^{-|g|}.
 \end{equation}
 That is, the lower tail is an exponential independent of the form of the replication distribution $p(k)$. 

The upper tail, as opposed to the lower tail, depends on the replication distribution $p(k)$ through $P_G$ and can be approximated as
 \[P_g(g\gg0)\sim\e^{g}P_G(\e^g).\]
   
As was discussed previously, for a power law replication process $p(k)$ with $\gamma<3$ the resulting distribution $P_G$ converges to a Levy stable distribution.   
The Levy stable distributions can be shown  \cite{Nolan09} to have an asymptotic power law decay 
\begin{equation}\label{levy_asymptotic}
L_{\gamma-1}^{\beta}(x;c,\mu)\sim x^{-\gamma}
\end{equation}
in the $x \to \infty$ limit. 
This suggests that the resulting logarithmic growth rate distribution $P_g$ will have an exponentially decaying upper tail 
\begin{equation}\label{upper_tail}
P_g(g\gg0)\sim e^{-(\gamma-1) g}.
\end{equation}
For distributions with a well defined mean, i.e. with $\gamma>2$, the upper tail (\ref{upper_tail}) always decays faster then the lower tail (\ref{lower_tail}), which means that 
the logarithmic growth rate distribution $P_g$ is always asymmetric with more weight in the lower tail.
This asymmetry in the logarithmic growth distribution was observed in many systems.

\section{Correlated replication factors}\label{sec:correlations}
 
The model, as defined thus far, assumes that the replication factors are uncorrelated, other than the correlation induced by the constraint on the total number of elements. 
We elaborate on the model by adding correlations between elements of the same group. However, we are not adding correlation between elements of different groups.  
This means that at each time step the replication factors of elements belonging to the same group will be correlated while factors of elements belonging to  different groups will be uncorrelated. 

We modify the the replication process as follows; at each moment in time an element is replicated according to the the following replication factor $\hkijt$ such that the number of elements in group $i$ at time $t+1$ is given by
\begin{equation}
N^i_{t+1}=\sum_j \hkijt.
\end{equation} 
The single element replication factor is comprised of two parts: one corresponding to the overall attractiveness of the group denoted as $\zeta$, which we term the common replication factor, and an individual replication factor $\kjt$, which we have defined and used previously (and in the main text). 
We define the new replication factor $\hkijt$  of element $j$ belonging to group $i$ at time $t$ as  
\begin{equation}
\hkijt=\kjt+\zit,
\end{equation}
where $\kjt$ is the uncorrelated individual replication factor, an i.i.d random variable drawn from $p(k)$ and $\zit$ is the attractiveness of group $i$ at time $t$, an i.i.d random variable and is drawn from a distribution $p_\zeta$. The replication factor $\zit$ is common to all elements in group $i$ at time $t$ and can be thought of as the attractiveness of the group in the sense that it is a common replication factor. 
The replication factors $\hkijt$ of the different elements are identically distributed but are now correlated with a correlation factor given by
\begin{equation}
\rho=\frac{Cov(\hkijt,\hat{k}^i_{j't})}{\sigma_{\hat{k}}^2}=\frac{\sigma_\zeta^2}{\sigma_k^2+\sigma_\zeta^2},
\end{equation}
where $\sigma_\zeta^2=Var[\zeta]$ and $\sigma_k=Var[k]$.
The correlation coefficient depends on the ratio of the variances and, assuming $\sigma_\zeta$ is non diverging, will vanish for distributions $p(k)$ with diverging second moments.     
It is important to note that since $\zit$ is an i.i.d random variable drawn separately for each group, i.e. the common factor is uncorrelated among different groups, the replication factors of elements belonging to different groups are uncorrelated $Cov(\hkijt,\hat{k}^{i'}_{jt})=0$.

\subsection{The resulting growth rate distribution $P_{\hat{G}}$}

Given that there are $N_t$ elements at time $t$ we write
\begin{equation}
 N^i_{t+1}=\sum_{j=1}^{N^i_t}\hkijt=\sum_{j=1}^{N^i_t}[\kjt+\zit]         
\end{equation} 
and the growth rate is given by
\begin{equation}
\hat{G}^i_t=\zit+G^i_t,
\end{equation}
where $G^i_t$ is the growth rate without correlations as discussed in the main text.  
The growth rate distribution $P_{\hat{G}}$ is given by 
\begin{equation}
P_{\hat{G}}(\hat{G}_t)=[P_G\ast P_\zeta](\hat{G}_t),
\end{equation}
which is the convolution of  the distribution $P_G$ and $P_\zeta$ and the tails of the resulting distribution  $P_{\hat{G}}$ are determined by the distribution with the heavier tail of the two.  
As the size of the group increases, $P_G$ will converge to a stable levy distribution. The speed of convergence and the parameters of the distribution depend on $p(k)$. 
If $P_\zeta$ has heavier tail then $P_G$, then the tails of $P_{\hat{G}}$ are determined by $P_\zeta$. However, the dominance of $P_\zeta$ over $P_G$ might occur only at large group sizes depending on the convergence of $P_G$. 

As an example for the behavior of $P_{\hat{G}}$, consider an individual replication distribution $p(k)\sim k^{-5}$ and a common replication factor distribution  $P(\zeta)\sim \exp(-\zeta)$. Since the replication distribution $p(k)$ has a finite second moment, we expect $P_G$ to converge, as the size of the group increases, to a normal distribution. However, this convergence is slow and for small group sizes, i.e. a small number of random summands comprising $G$, the distribution has not yet converged to a normal and will obey $P_G\sim G^{-5}$. 
The question is how will the distribution $P_{\hat{G}}$ behave? For small group sizes the distribution $P_G\sim G^{-5}$ dominates over $P_\zeta$ and we can expect the growth rate distribution to be $P_{\hat{G}}\sim\hat{G}^{-5}$. However, for large enough group sizes $P_G$ will have converged to a normal and will be dominated by $P_\zeta$ such that $P_{\hat{G}}\sim\exp(-\hat{G})$. Thus, as the group size increases, we observe a change in $P_{\hat{G}}$ such that for small groups the distribution obeys $P_{\hat{G}}\sim P_G$ while for large groups we will observe $P_{\hat{G}}\sim P_\zeta$.
The group size for which this transition occurs depends on the individual replication distribution $p(k)$ and the common replication factor distribution $P_\zeta$.

\subsection{The resulting fluctuation scaling of $\sigma_{\hat{G}}$ }
 
\begin{table}
\begin{center}
\begin{tabular}{c|c}
\hline
Degree Distribution &$\beta$\\
\hline\hline
$p(k)\sim k^{-\gamma}$ with $1<\gamma<3$ &$\frac{\gamma-2}{\gamma-1}$\\
\hline
$p(k)\sim k^{-\gamma}$ with $3<\gamma$ &$\frac{1}{2}$\\
\hline
Thinner tail than a power law &$\frac{1}{2}$\\
\hline
\end{tabular}
\caption{The dependence of the relative growth rate fluctuation scaling exponent $\beta$ on the underlying replication distributions $p(k)$. 
\label{table_scaling}}
\end{center}
\end{table}%

Since the individual replication factor $\kijt$ is independent of the common factor $\zit$, the relative growth rate variance $\sigma_{\hat{G}}$ is given by
\begin{equation}
\sigma_{\hat{G}}^2=\sigma_\zeta^2+\sigma_G^2,
\end{equation}
where $\sigma_G$ depends on size according to the fluctuation scaling $\sigma_G=\sigma_1 N^{-\beta}$ with $\beta$ as given in Table~\ref{table_scaling}.
The fluctuation scaling  as a function of the group size is given by
\begin{equation}\label{sigma_hat_G}
\sigma_{\hat{G}}^2=\sigma_\zeta^2+\sigma_1^2N^{-2\beta}.
\end{equation}

For $\sigma_\zeta<\sigma_1$, the second term in the RHS of equation (\ref{sigma_hat_G}) dominates for small $N$ while for large enough $N$ the first term will dominate. 
We define the transition size $N^*$ as the group size for which the volatility of both terms on the RHS are equal such that
\[
N^*=\left(\frac{\sigma_1}{\sigma_\zeta}\right)^{\frac{2}{\beta}}
\]
and the volatility $\sigma_{\hat{G}}$ obeys
\begin{eqnarray}
N\ll N^* \,\,\,& \sigma_{\hat{G}}&= \sigma_1N^{-\beta}\nonumber\\
N\gg N^*\,\,\,&\sigma_{\hat{G}}&= \sigma_\zeta.
\end{eqnarray}
This corresponds to a fluctuation scaling exponent $\hat{\beta}$ that depends on size such that 
\begin{eqnarray}
N\ll N^* \,\,\,& \hat{\beta}&= \beta\nonumber\\
N\gg N^*\,\,\,&\hat{\beta}&= 0.
\end{eqnarray}
For $\sigma_\zeta\geq\sigma_1$ there is no transition size and the volatility can be approximated as
 \[%
\sigma_{\hat{G}}\approx \sigma_\zeta.
\]%
and as a result the fluctuations are independent of size, i.e. $\hat{\beta}=0$, in accordance with Gibrat's law.

We can conclude that given a replication process with a common replication factor, for large $N$ the fluctuations obey Gibrat's law. However, the size of the group for which this transition takes place can be very large such that effectively we can have $\hat{\beta}= \beta$ for the entire observation range.  

\section{Empirical evidence}\label{sec:empirics}

In this section we describe in more detail the empirical evidence for three data sets; The growth rate of breeding birds of North America in Section~\ref{sec:birds}, the growth, due to investors, of mutual funds for the years 1997 to 2007 in Section~\ref{sec:mf_data} and the growth of firms with respect to sales in Section~\ref{sec:firm_data}. 
 \subsection{Methods}
 The empirical investigation of the three data sets was conducted as follows:
 first, the fluctuation scaling exponent $\beta$ is estimated from the data. Then, we normalize the growth rate distribution followed by an estimation of the tail exponent $\gamma$. Lastly we will compare the measured $\beta$ and $\gamma$ to the expected relationship from our model.
To estimate the fluctuation scaling exponent $\beta$ the relative growth rate distribution $G=N_{t+1}/N_t-1$ was binned into 20 exponentially spaced bins according to size $N_t$. For each bin $i$, the variance of the growth rates  $\sigma_i^2$ was empirically estimated.
Then the logarithm of the measured variances were regressed on the logarithm of the average size $\bar{N}_i$
\begin{equation}
\log(\sigma)=\beta \log(N)+\sigma_1
\end{equation}
 such that the slope is the ordinary least squares (OLS) estimator of $\beta$.
 Using the fluctuation scaling exponent $\beta$ we predict the tail exponent 
\[\hat{\gamma}=\frac{2-\beta}{1-\beta},\]
which we will compare to the measured maximum likelihood estimator (MLE) of the tail exponent. 
 To estimate the tail exponent, we normalize the growth rate $G$ such that it has zero mean and unit variance.  For distributions with a power law tail, the MLE power law exponent was estimated using the technique described in Clauset et al \cite{Clauset07} for fitting power law tails. The method used uses the following modified Kolmogorov-Smirnoff statistic 
\[KS=\max_{x>x_{min}}\frac{|s(x)-p(x)|}{\sqrt{p(x)[1-p(x)]}},\] 
where $s$ is the empirical cumulative distribution and $p$ is the hypothesized cumulative distribution.

\subsection{Breeding birds of America}\label{sec:birds}

We use the the North American breeding bird survey\footnote{The data set can be found on line at 
ftp://ftpext.usgs.gov/pub/er/md/laurel/BBS/DataFiles/.} %
which contains 42 yearly abundance observations for over 600 species along more than 3,000 observation routes.
 For each route the number of birds from each specie is quoted for each year in the period 1966-2007.

\begin{figure}
\begin{center}
\includegraphics[width=7cm]{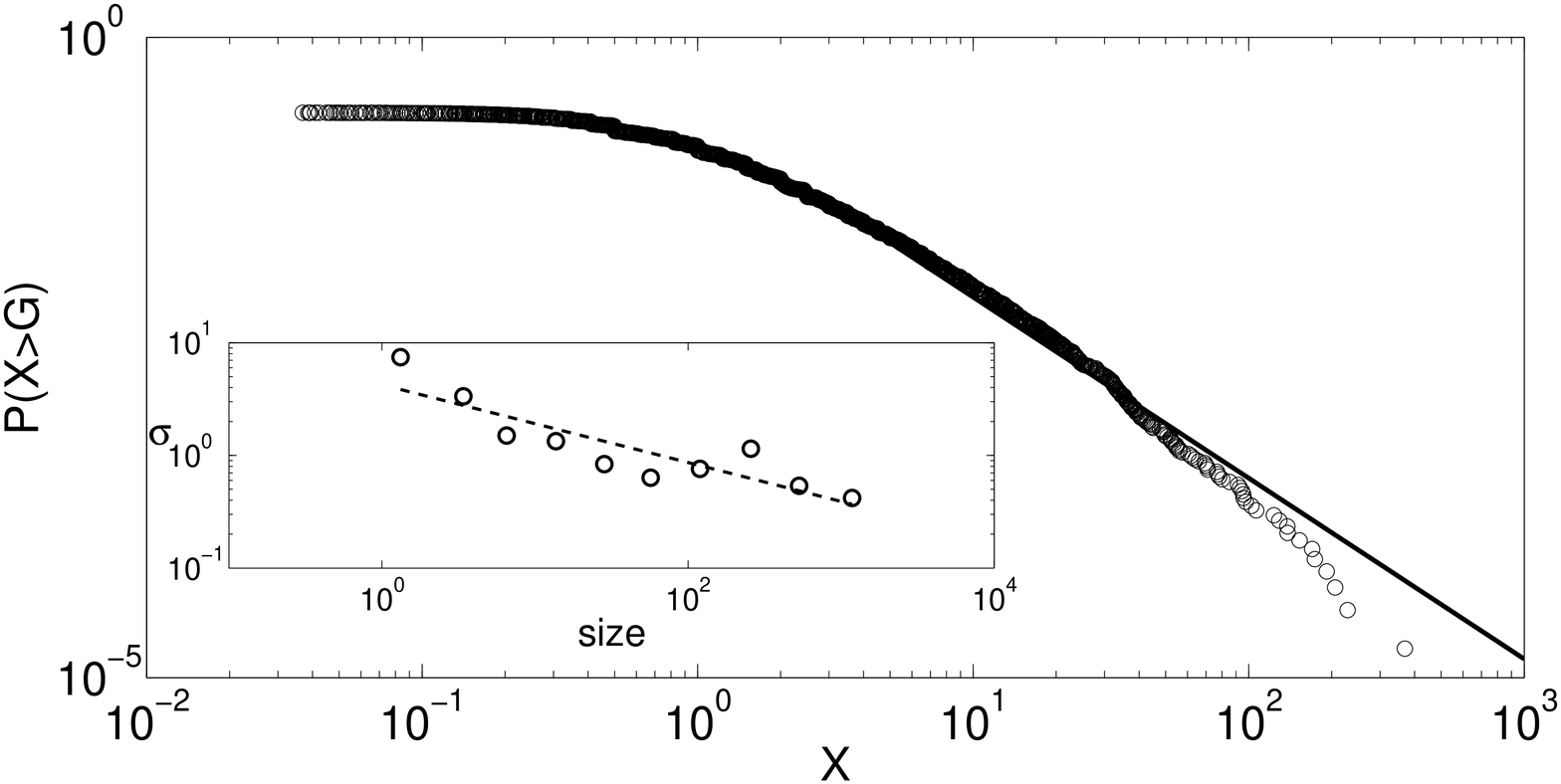}
\includegraphics[width=7cm]{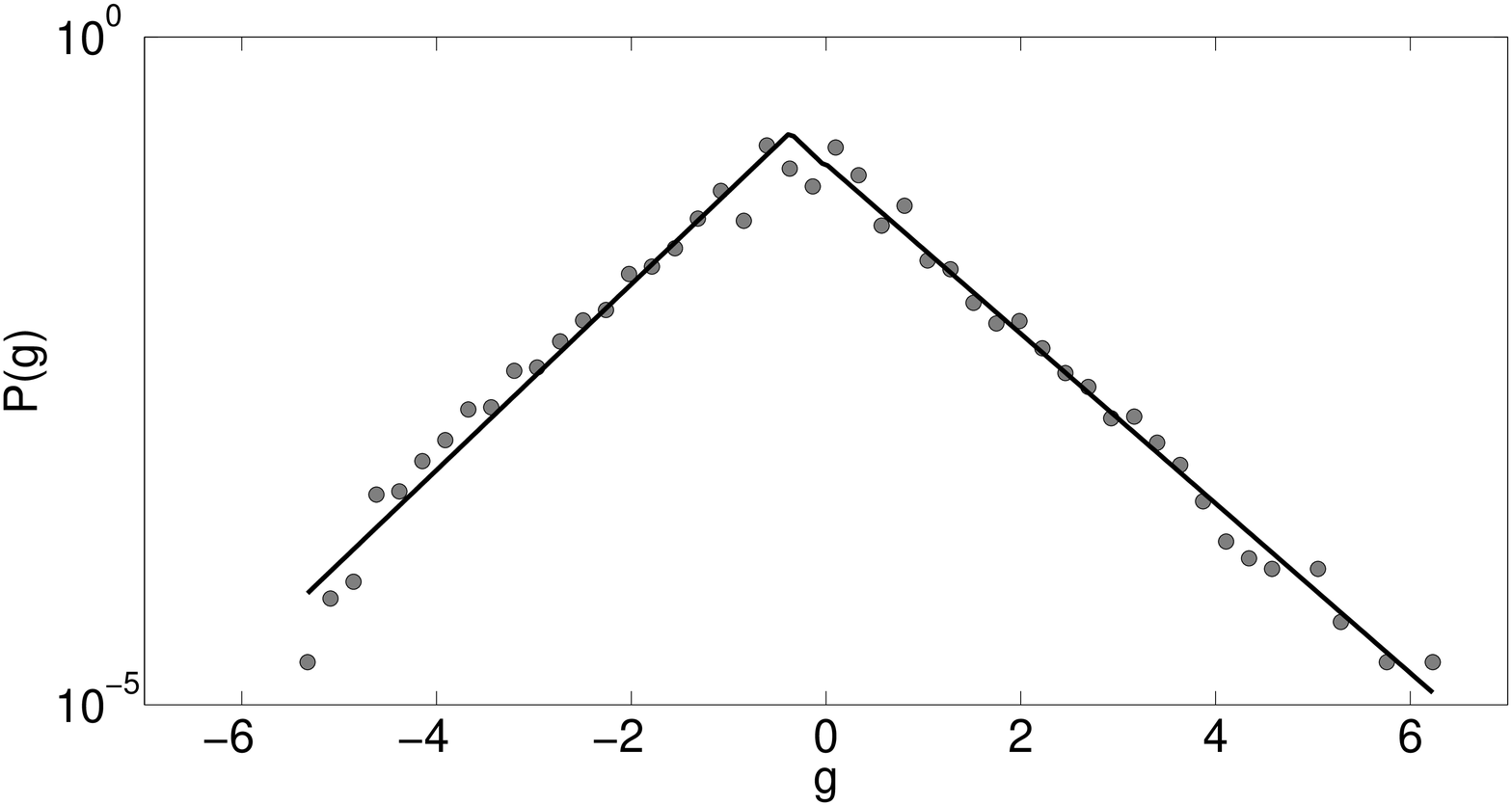}
\caption{Left panel: The cumulative distribution for the relative growth rates $P(G>X)$ for the year 2007 is plotted on a log-log scale. The distributions is compared to a linear line with a lope corresponding to the upper tail exponent from the MLE fit for $\gamma$ given in Table~\ref{table_birds_fit}. 
Inset:  The relative growth rate fluctuations $\sigma$ for the year 2007 as a function of the number of birds, measured as yearly sales, is compared to a line with a slope corresponding to $\beta$ given in Table~\ref{table_birds_fit}. \newline
Right panel: The relative growth rates density $P_g(g)$ for the year 2007, as resulting form binning the data, is plotted on a semi-log scale. The distributions is compared to a laplacian distribution.}
\label{birds_pg}
\end{center}
\end{figure}

\begin{table}
\tiny
\begin{center}
\begin{tabular}{|c|c|c|c|}
\hline
year&$\beta$&$\hat{\gamma}$&$\gamma$\\
\hline\hline
$1966$&$0.39\pm0.10$&$2.64\pm0.27$&$2.37\pm0.11$\\
\hline
$1967$&$0.36\pm0.10$&$2.56\pm0.25$&$2.53\pm0.13$\\
\hline
$1968$&$0.44\pm0.15$&$2.77\pm0.46$&$2.39\pm0.09$\\
\hline
$1969$&$0.30\pm0.15$&$2.43\pm0.30$&$2.45\pm0.12$\\
\hline
$1970$&$0.42\pm0.11$&$2.72\pm0.31$&$2.29\pm0.08$\\
\hline
$1971$&$0.31\pm0.15$&$2.44\pm0.32$&$2.38\pm0.11$\\
\hline
$1972$&$0.30\pm0.11$&$2.44\pm0.23$&$2.61\pm0.21$\\
\hline
$1973$&$0.32\pm0.13$&$2.47\pm0.27$&$2.53\pm0.13$\\
\hline
$1974$&$0.26\pm0.15$&$2.35\pm0.28$&$2.61\pm0.17$\\
\hline
$1975$&$0.33\pm0.05$&$2.48\pm0.12$&$2.64\pm0.10$\\
\hline
$1976$&$0.33\pm0.10$&$2.49\pm0.21$&$2.46\pm0.10$\\
\hline
$1977$&$0.29\pm0.11$&$2.41\pm0.22$&$2.65\pm0.14$\\
\hline
$1978$&$0.44\pm0.12$&$2.78\pm0.37$&$2.42\pm0.12$\\
\hline
$1979$&$0.50\pm0.62$&$3.00\pm2.48$&$2.63\pm0.14$\\
\hline
$1980$&$0.41\pm0.14$&$2.70\pm0.41$&$2.48\pm0.10$\\
\hline
$1981$&$0.30\pm0.09$&$2.42\pm0.19$&$2.43\pm0.08$\\
\hline
$1982$&$0.40\pm0.12$&$2.66\pm0.32$&$2.85\pm0.27$\\
\hline
$1983$&$0.43\pm0.13$&$2.74\pm0.40$&$2.58\pm0.15$\\
\hline
$1984$&$0.34\pm0.14$&$2.52\pm0.33$&$2.31\pm0.07$\\
\hline
$1985$&$0.33\pm0.18$&$2.49\pm0.40$&$2.36\pm0.10$\\
\hline
$1986$&$0.34\pm0.10$&$2.51\pm0.24$&$2.46\pm0.12$\\
\hline
$1987$&$0.35\pm0.19$&$2.55\pm0.45$&$2.26\pm0.05$\\
\hline
$1988$&$0.32\pm0.13$&$2.47\pm0.28$&$2.60\pm0.13$\\
\hline
$1989$&$0.33\pm0.06$&$2.49\pm0.12$&$2.33\pm0.06$\\
\hline
$1990$&$0.32\pm0.12$&$2.47\pm0.26$&$2.76\pm0.18$\\
\hline
$1991$&$0.30\pm0.15$&$2.43\pm0.30$&$2.73\pm0.17$\\
\hline
$1992$&$0.35\pm0.08$&$2.54\pm0.19$&$2.52\pm0.08$\\
\hline
$1993$&$0.35\pm0.14$&$2.55\pm0.33$&$2.87\pm0.16$\\
\hline
$1994$&$0.34\pm0.11$&$2.53\pm0.26$&$2.53\pm0.11$\\
\hline
$1995$&$0.31\pm0.11$&$2.46\pm0.24$&$2.42\pm0.06$\\
\hline
$1996$&$0.48\pm0.26$&$2.91\pm0.96$&$2.71\pm0.12$\\
\hline
$1997$&$0.40\pm0.26$&$2.66\pm0.71$&$2.35\pm0.05$\\
\hline
$1998$&$0.30\pm0.10$&$2.43\pm0.20$&$2.61\pm0.10$\\
\hline
$1999$&$0.34\pm0.10$&$2.51\pm0.24$&$3.06\pm0.25$\\
\hline
$2000$&$0.22\pm0.18$&$2.29\pm0.30$&$2.51\pm0.09$\\
\hline
$2001$&$0.32\pm0.16$&$2.48\pm0.36$&$2.69\pm0.10$\\
\hline
$2002$&$0.39\pm0.11$&$2.65\pm0.29$&$2.70\pm0.11$\\
\hline
$2003$&$0.42\pm0.22$&$2.71\pm0.65$&$2.71\pm0.14$\\
\hline
$2004$&$0.30\pm0.14$&$2.42\pm0.28$&$2.55\pm0.08$\\
\hline
$2005$&$0.27\pm0.17$&$2.37\pm0.33$&$2.95\pm0.17$\\
\hline
$2006$&$0.33\pm0.08$&$2.49\pm0.18$&$2.54\pm0.09$\\
\hline
$2007$&$0.30\pm0.14$&$2.43\pm0.29$&$2.41\pm0.06$\\
\hline
mean&$0.35\pm0.02$&$2.54\pm0.04$&$2.55\pm0.01$\\
\hline
\end{tabular}
\caption{\label{table_birds_fit}%
\small The parameter values for the proposed growth model as measured for from the north american breeding bird data.
The parameters of the model were measured for the aggregated monthly data during each year and the errors are the 95\% 
confidence interval.
The last row corresponds to the average value over the different years.
The parameters are as follows:\newline
$\beta$ - The OLS estimator for the exponent of the size dependence of the fluctuations in the relative growth rate. \newline
$\hat{\gamma}$ - The predicted power law exponent as can be inferred from $\beta$. To be compared with $\gamma$. \newline
$\gamma$ - MLE for the power law exponent of the upper tail of $P(G)$.}
\end{center}
\end{table}

For each year in the data set, from 1966 to 2007, we computed the yearly growth with respect to each specie in each route and compared the data to the model.
For each year  we computed the Maximum Likelihood Estimators (MLE) for the power law exponents of the upper tail. The results are summarized in Table~\ref{table_birds_fit} and the growth rate distribution for the year 2007 is plotted in Figure~\ref{birds_pg}. 
Remarkably, for most years, the MLE of the upper law tail exponent  $\gamma$ is in agreement with the estimations from the growth fluctuations $\hat{\gamma}$ . 
Moreover, for each of these values the mean over the different years was calculated and all three are in agreement. 
Our model seems to describe well the relationship between the growth rate distribution 
and the growth rate fluctuations observed observed the North American breeding birds.

\subsection{Mutual Funds}\label{sec:mf_data}

In this section we investigate growth in the mutual fund industry and compare it to the model using  the CRSP mutual fund data base. 
We restrict ourselves to equity mutual funds existing in the years 1997 to 2007.
We defined equity funds as funds with at least 80\% of their portfolio in stocks and cash.
As the size of the Mutual fund we use the total net assets value (TNA) in real US dollars as  reported monthly in the data base.  
Growth in the mutual fund industry, measured by change in TNA, is comprised of two sources: growth due to the funds performance and growth due to flux of money from investors, i.e. mutual funds can grow in size if their assets increase in value or due to new money coming in from investors.
We define the relative growth in the size of a fund at time $t$ as 
\[G_{TNA}(t)=\frac{TNA_{t+1}}{TNA_t}-1\]
and decompose it as follows;
\begin{equation}
G_{TNA}(t)=r_t+G_t,
\end{equation} 
where $r_t$ is the fund's return, quoted monthly in the database, and $G_t$ is the growth due to investors.  
Since the growth due to the fund's performance is not part of our model, we restrict ourselves to growth due to investors either investing money in the funds or withdrawing money from the funds. 

\begin{table}
\begin{center}
\begin{tabular}{|c||c|c|c|}
\hline
year&$\beta$&$\hat{\gamma}$&$\gamma$\\
\hline\hline
$1997$&$0.34\pm0.11$&$2.52\pm0.25$&$2.36\pm0.12$\\
\hline
$1998$&$0.38\pm0.09$&$2.62\pm0.24$&$2.67\pm0.17$\\
\hline
$1999$&$0.49\pm0.08$&$2.98\pm0.30$&$2.53\pm0.09$\\
\hline
$2000$&$0.35\pm0.04$&$2.55\pm0.09$&$2.58\pm0.08$\\
\hline
$2001$&$0.38\pm0.06$&$2.62\pm0.17$&$2.85\pm0.14$\\
\hline
$2002$&$0.44\pm0.08$&$2.80\pm0.26$&$2.50\pm0.08$\\
\hline
$2003$&$0.32\pm0.05$&$2.47\pm0.10$&$2.68\pm0.11$\\
\hline
$2004$&$0.37\pm0.07$&$2.60\pm0.17$&$2.63\pm0.09$\\
\hline
$2005$&$0.42\pm0.10$&$2.71\pm0.29$&$2.35\pm0.07$\\
\hline
$2006$&$0.34\pm0.06$&$2.52\pm0.15$&$2.56\pm0.08$\\
\hline
$2007$&$0.29\pm0.08$&$2.42\pm0.16$&$2.53\pm0.08$\\
\hline
mean&$0.38\pm0.02$&$2.62\pm0.05$&$2.57\pm0.03$\\
\hline
\end{tabular}
\end{center}
\caption{\small The parameter values for the proposed growth model as measured for monthly size growth (in US dollars)
 of equity mutual funds using the CRSP data base.
The parameters of the model were measured for the aggregated monthly data during each year and the errors are the 95\% 
confidence interval.
The last row corresponds to the average value over the different years.
The parameters are as follows: \newline
$\beta$ - The OLS estimator for the exponent of the size dependence of the fluctuations in the relative growth rate. \newline
$\hat{\gamma}$ - The predicted power law exponent as can be inferred from $\beta$. To be compared with $\gamma$. \newline
$\gamma$ - MLE for the power law exponent of the upper tail of $P(G)$.}
\label{mutual_fund_fits}
\end{table}%

\begin{figure}
\begin{center}
\includegraphics[width=7cm]{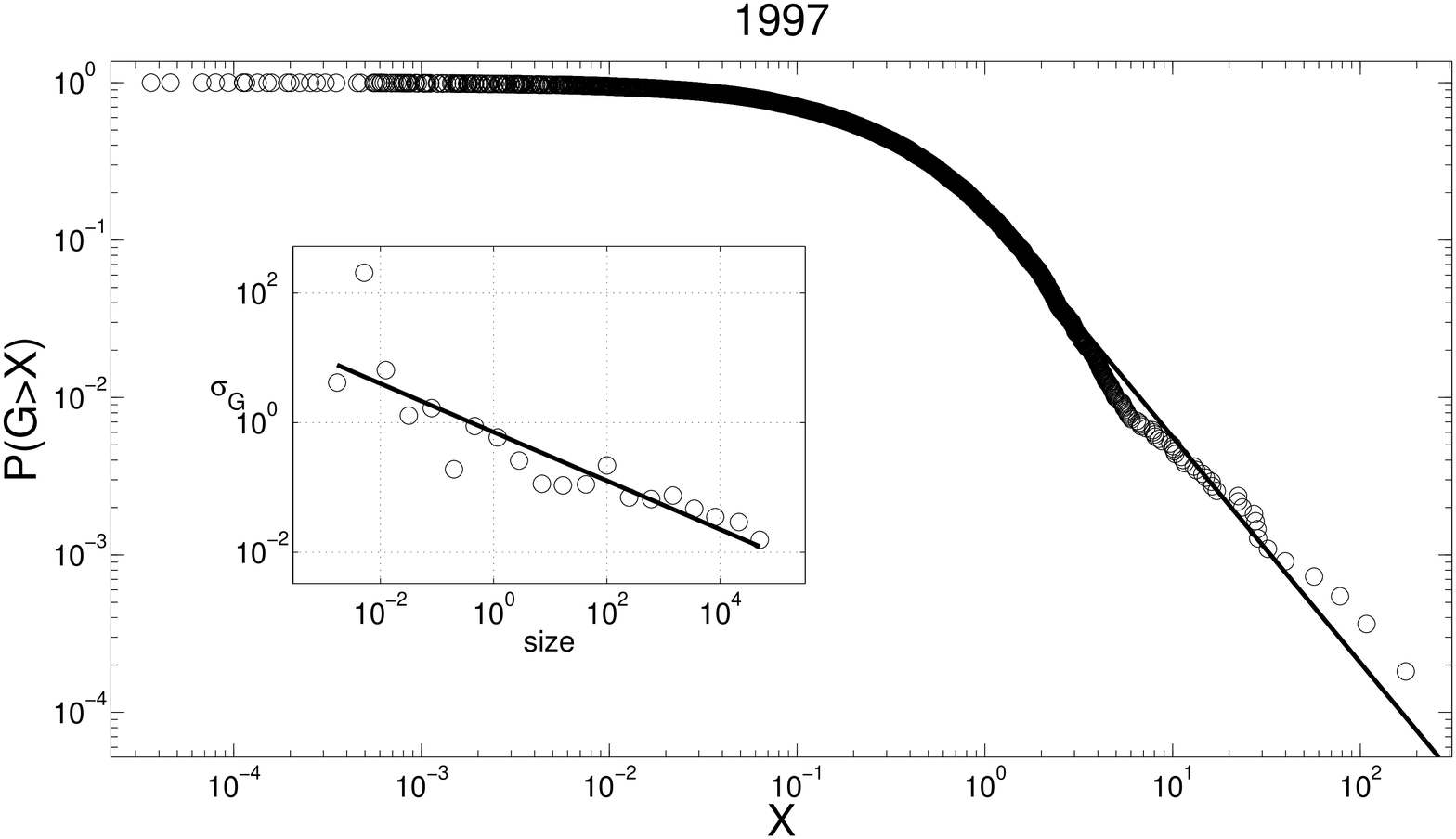}
\includegraphics[width=7cm]{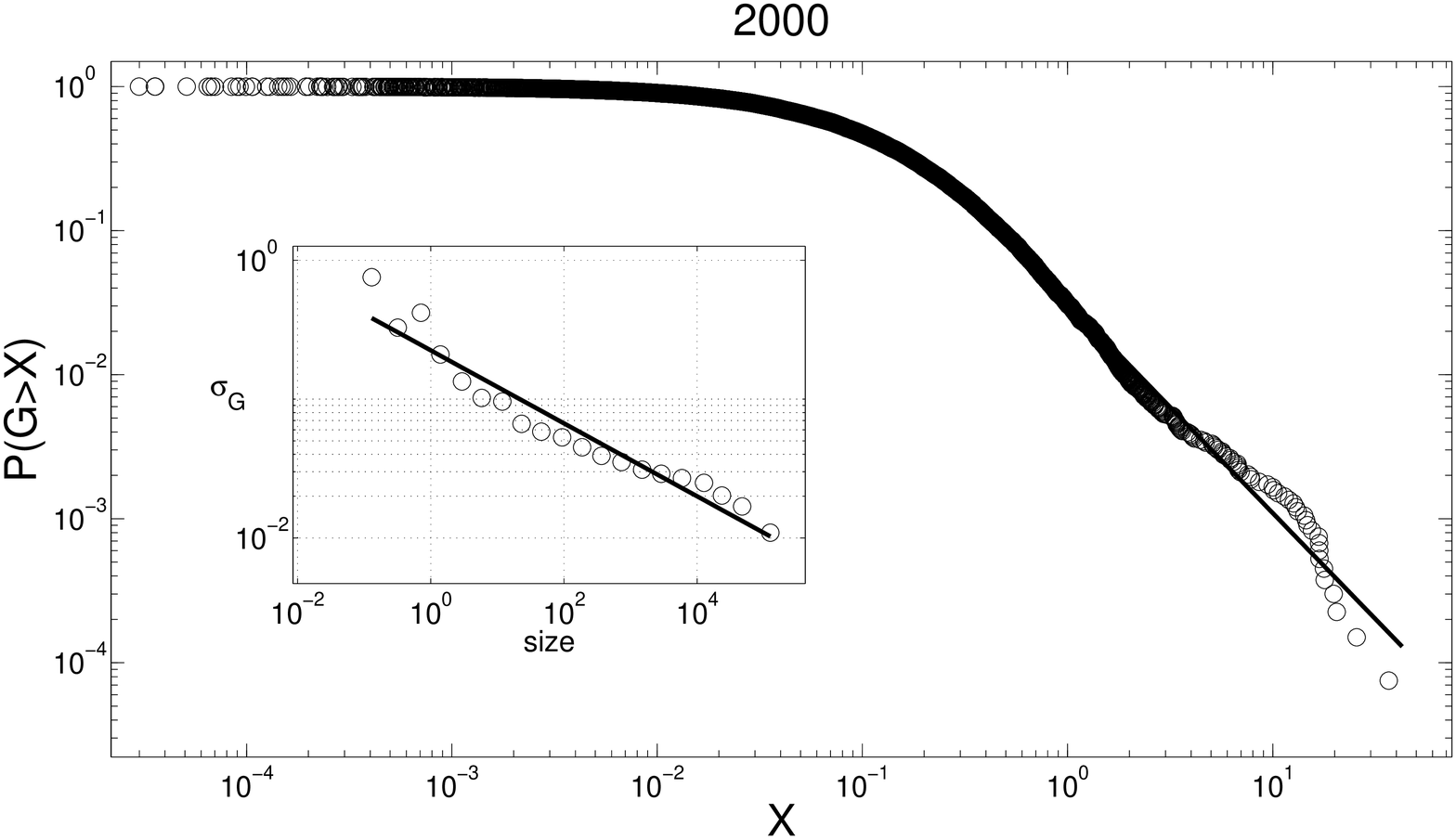}
\includegraphics[width=7cm]{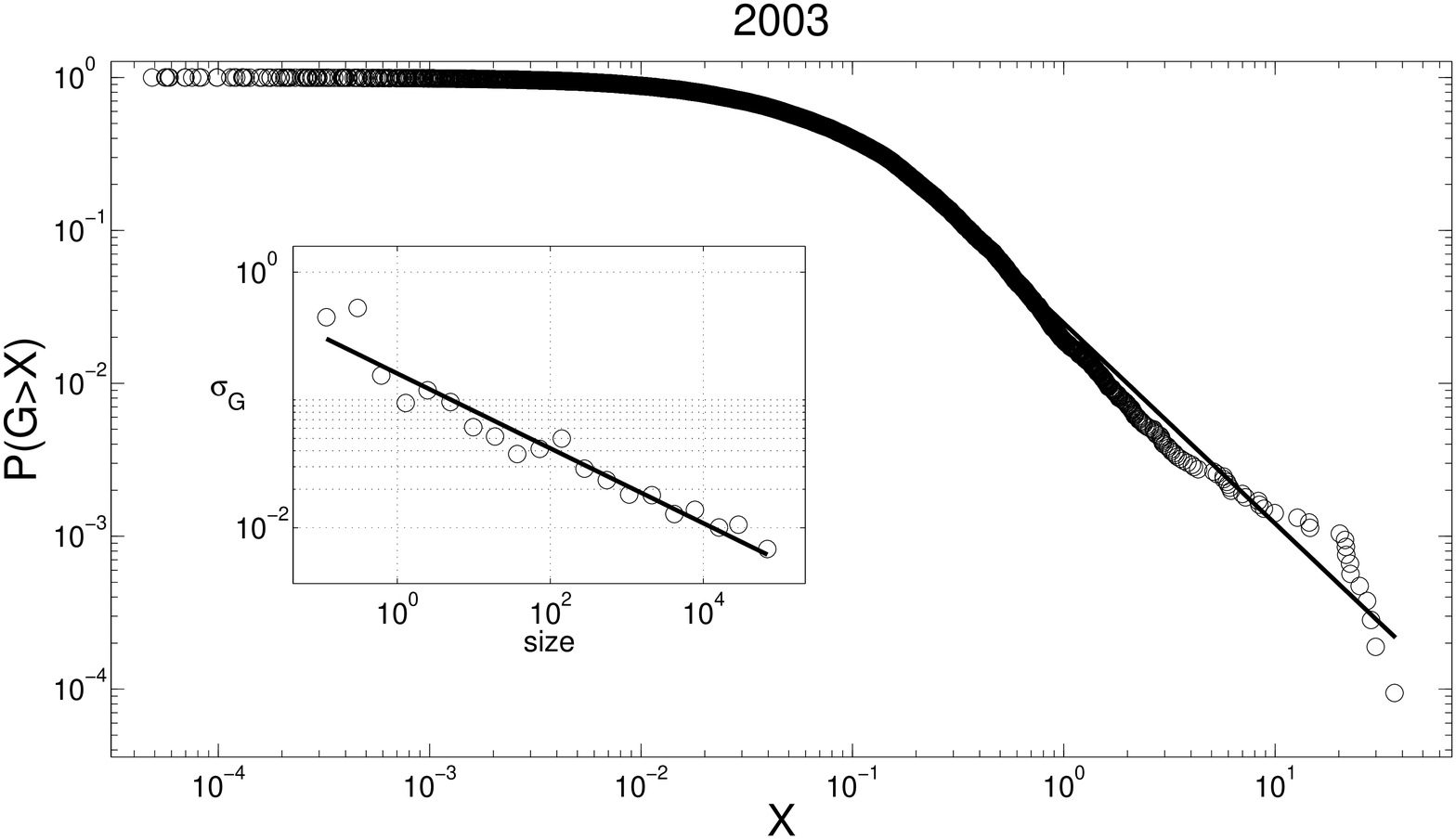}
\includegraphics[width=7cm]{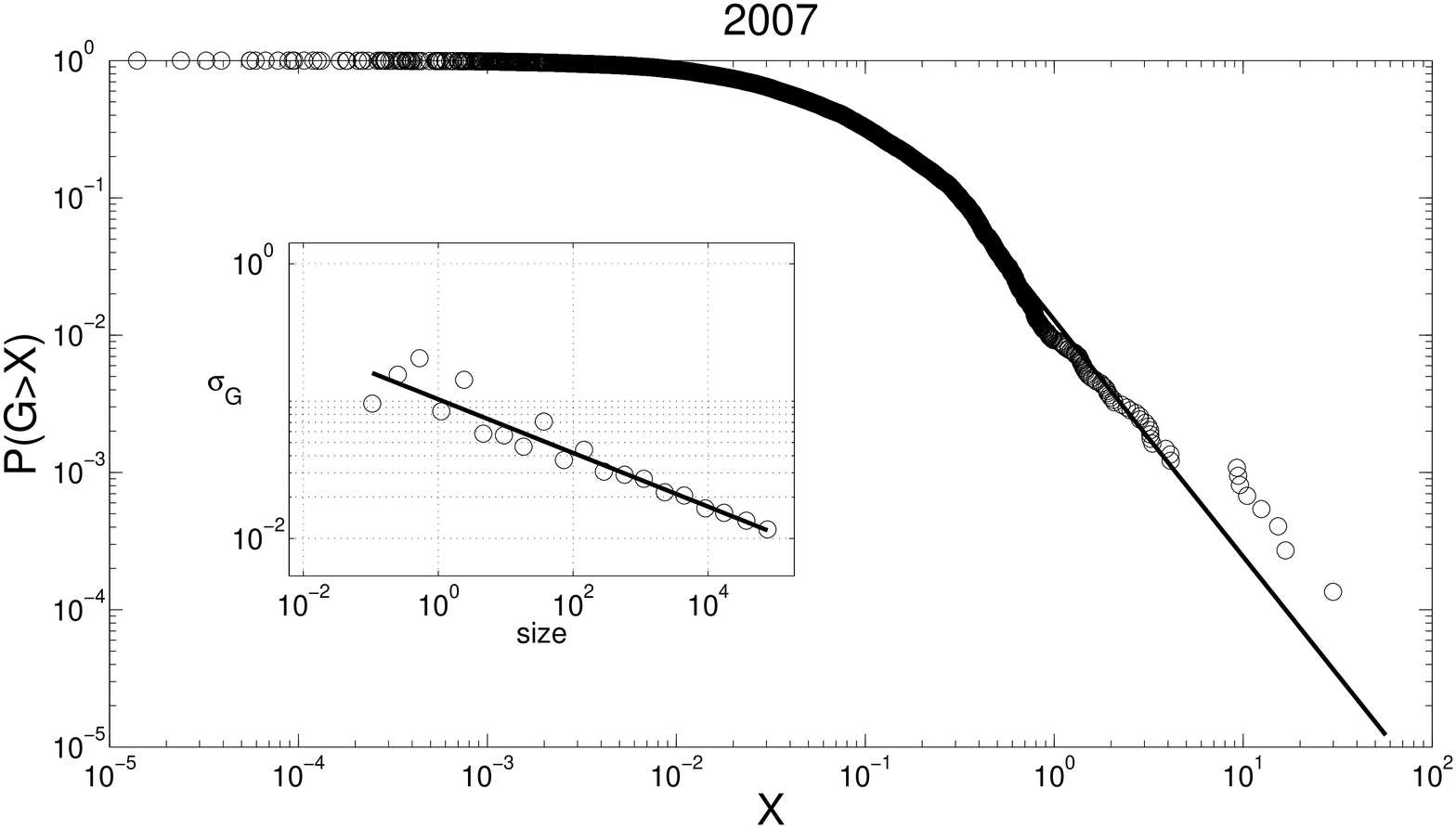}
\caption{The cumulative distribution for the relative growth rates $P(G>X)$ are plotted on a log-log scale for the years 1997, 2000, 2003 and 2007. The upper tail is compared to a line on a log-log plot with a slope corresponding to the MLE fits for $\gamma$ given in Table~\ref{mutual_fund_fits}. 
Insets:  The relative growth rate fluctuations $\sigma$ as a function of the size of the mutual fund are compared to lines with slopes corresponding to $\beta_G$ given in Table~\ref{mutual_fund_fits}. }
\label{mf_pg}
\end{center}
\end{figure}

 For each year  we computed the Maximum Likelihood Estimators (MLE) for the power law exponents of the upper tail. The results are summarized in Table~\ref{mutual_fund_fits} and the fits for some of the years are given in Figure~\ref{mf_pg}. 
For all years, the upper law tail exponent is in the range $\gamma\in(2,3)$ and except for the years 1999,2002 and 2005 the measured value $\gamma$ is in agreement with the estimations from the growth fluctuations $\hat{\gamma}$. 
Moreover, for each of these values the mean over the different years was calculated and all three are in agreement. 
Our model seems to describe well the relationship between the growth rate distribution 
and the growth rate fluctuations observed in the mutual fund industry. 

Another interesting and non trivial observation is that the tail is heavy in the sense that the second moment does not exist especially when one considers the fact that this growth is solely due to investors. It seems that in mutual funds the individual replication dominates over the common replication and growth is mostly affected by the ability of each new client to generate more clients than the mere attractiveness of the fund.

\subsection{Firm Growth}\label{sec:firm_data}
Using the 2008 COMPUSTAT data set we test our model on the growth of public firms. 
As the size of a firm we use the dollar amount of sales and growth is given by the growth in 3 year sales. 
The OLS estimator for $\beta$, the resulting tail exponent prediction $\hat{\gamma}$ and the MLE tail exponent $\gamma$
are summarized in Table~\ref{firm_fit}.
Our model seems to describe well the relationship between the growth rate distribution 
and the growth rate fluctuations observed for the growth of public firms.
\begin{table}
\begin{center}
\begin{tabular}{|c|c|c|}
\hline
$\beta$&$\hat{\gamma}$&$\gamma$\\
\hline\hline
$0.31\pm0.07$&$2.45\pm0.15$&$2.53\pm0.07$\\
\hline
\end{tabular}
\end{center}
\caption{\small The parameter values for the proposed growth model as measured for the yearly growth (in US dollars)
 of firm sales using the COMPUSTAT data base.
The errors are the 95\%  confidence interval.
The parameters are as follows: \newline
$\beta$ - The OLS estimator for the exponent of the size dependence of the fluctuations in the relative growth rate. \newline
$\hat{\gamma}$ - The predicted power law exponent as can be inferred from $\beta$. To be compared with $\gamma$. \newline
$\gamma$ - MLE for the power law exponent of the upper tail of $P(G)$.}
\label{firm_fit}
\end{table}%
\begin{figure}
\begin{center}
\includegraphics[width=8cm]{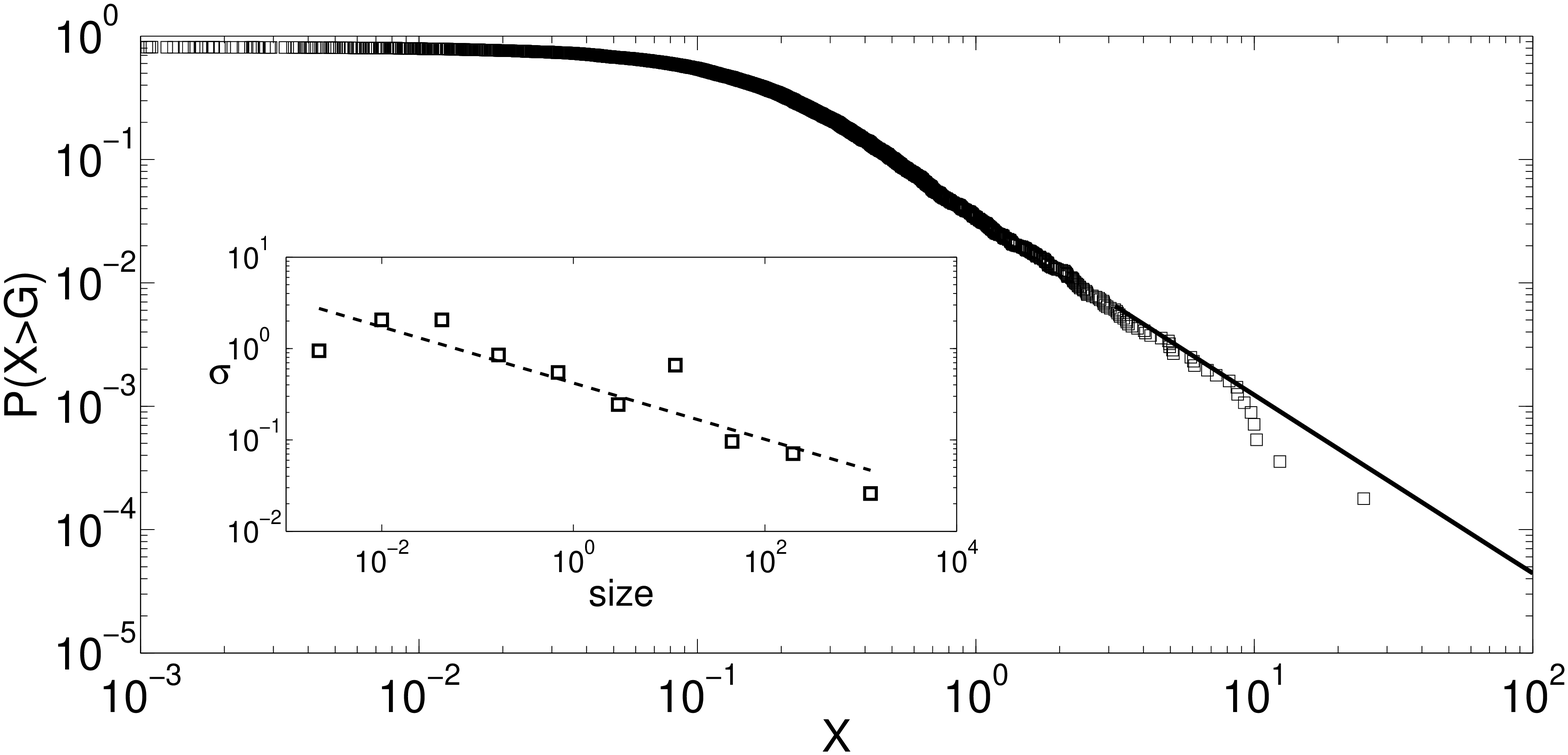}
\includegraphics[width=8cm]{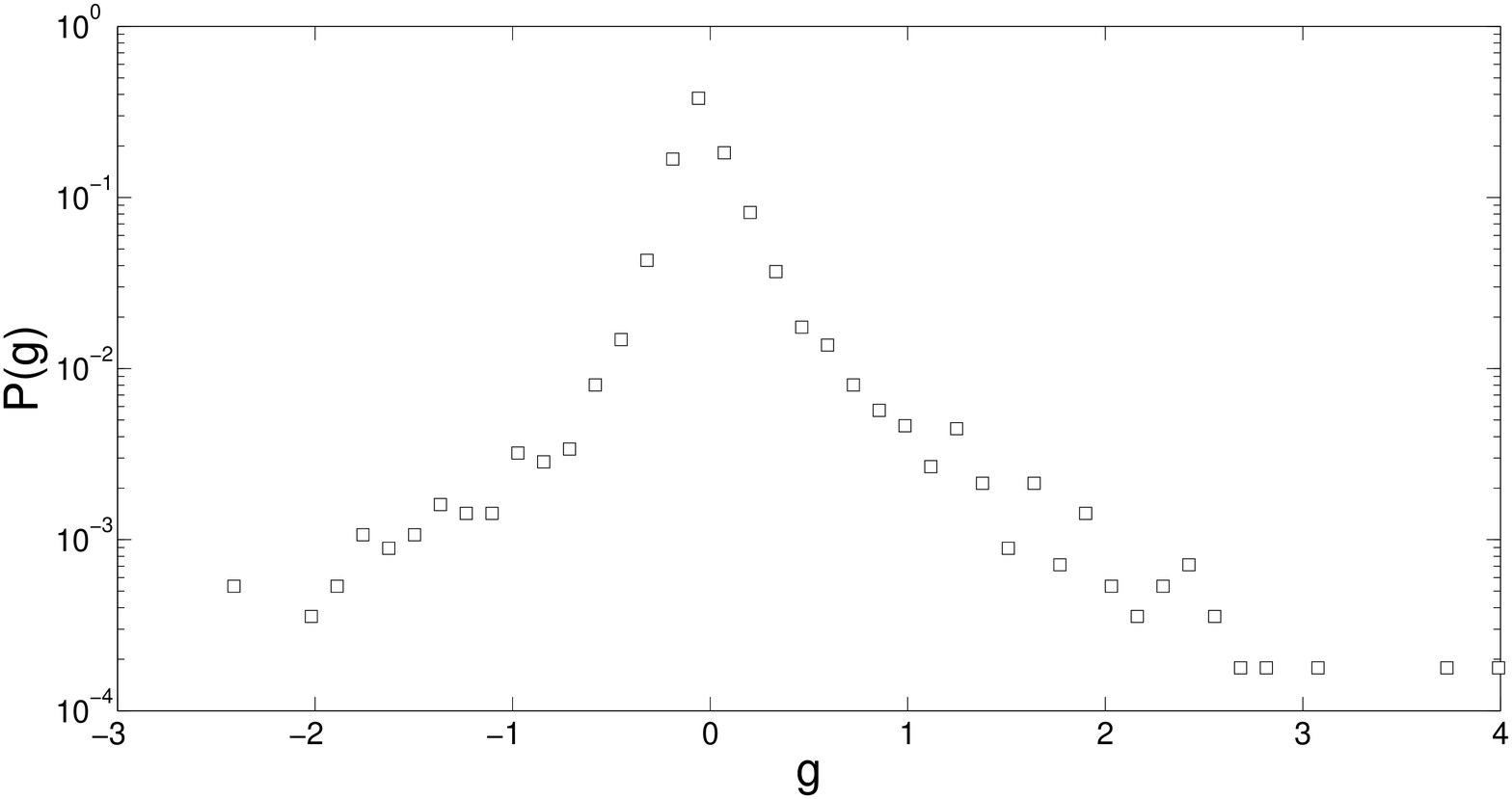}
\caption{Left panel: The cumulative distribution for the relative growth rates $P(G>X)$ is plotted on a log-log scale. The upper tail is compared to a line on a log-log plot with a slope corresponding to the MLE fits for $\gamma$ given in Table~\ref{firm_fit}.
Inset:  The relative growth rate fluctuations $\sigma$ as a function of the size of the company, measured as yearly sales, is compared to a line with a slope corresponding to $\beta$ given in Table~\ref{firm_fit}. \newline
Right panel: The logarithmic growth rate distribution for the relative growth rates $P_g(g)$, as resulting from binning the data, is plotted on a semi-log scale.}
\label{firm_pg}
\end{center}
\end{figure}

\bibliographystyle{plain}
\bibliography{growth}